\documentclass[preprint2]{aastex}

\newcommand{\kms}{km~s$^{-1}$~}

\newcommand{\mbhg}{\ensuremath{M_\mathrm{BH}}}
\newcommand{\lf}{\ensuremath{L_{\rm{5100 \AA}}}}
\newcommand{\fwha}{\ensuremath{\mathrm{FWHM}_\mathrm{H{\alpha}}}}
\def\mbh{{$\mathcal M_{\rm BH}$}}
\def\sig{{$\sigma_*$}}
\def\msol{{$\mathcal M_{\odot}$}}

\def\msig{{$\mathcal M_{\rm BH}$-$\sigma_*$}}
\def\mbulge{{$\mathcal M_{\rm bulge}$}}

\def\ha{{H$\alpha$}}

\shorttitle{Probing \msig\ using Red QSOs}
\shortauthors{Canalizo et al.}

\begin{document}

\title{Probing the \msig\ relation in the non-local universe using red QSOs}

\author{Gabriela Canalizo\altaffilmark{1}, Margrethe Wold\altaffilmark{2}, Kyle D. Hiner\altaffilmark{1},
Mariana Lazarova\altaffilmark{1},
Mark Lacy\altaffilmark{3}, Kevin Aylor\altaffilmark{1}}

\altaffiltext{1}{Department of Physics and Astronomy, 
University of California, Riverside, CA 92521, USA;
email: gabriela.canalizo@ucr.edu, kyle.hiner@student.ucr.edu, mariana.lazarova@student.ucr.edu, kevin.aylor@email.ucr.edu}

\altaffiltext{2}{Dark Cosmology Centre, Niels Bohr Institute, University of Copenhagen, Juliane Maries Vej 30, DK-2100, Denmark; email: mwold@dark-cosmology.dk}

\altaffiltext{3}{North American ALMA Science Center, National Radio Astronomy Observatory, Charlottesville, VA 22903, USA; email: mlacy@nrao.edu}

\begin{abstract}
We describe a method to measure the \msig\ relation in the non-local universe using dust-obscured QSOs.   
We present results from a pilot sample of nine 2MASS red QSOs with redshifts $0.14<z<0.37$.  We find that there is an offset (0.8 dex, on average) between the position of our objects and the local 
relation for AGN, in the sense that the majority of red QSO hosts have lower velocity dispersions and/or more massive BHs than local galaxies.   These results are in agreement with recent studies of AGN at similar and higher redshifts.  This could indicate an unusually rapid growth in the host galaxies since $z\sim0.2$, if these objects were to land in the local relation at present time.   However, the $z>0.1$ AGN (including our sample and those of previous studies) have significantly higher \mbh\ than those of local AGN, so a direct comparison is not straightforward.   Further, using several samples of local and higher-$z$ AGN, we find a striking trend of an increasing offset with respect to the local \msig\ relation as a function of AGN luminosity, with virtually all objects with log(L$_{5100}$/erg s$^{-1}$) $> 43.6$ falling above the relation. 
Given the relatively small number of AGN at $z>0.1$ for which there are direct measurements of stellar velocity dispersions, it is impossible at present to determine whether there truly is evolution in \msig\ with redshift.  Larger, carefully selected samples of AGN are necessary to disentangle the dependence of \msig\ on mass, luminosity, accretion rates, and redshift. 
\end{abstract}


\section{Introduction}

During the last decade, the astronomy community has come to realize that black hole (BH) activity plays an important role in galaxy formation, e.g., by regulating star formation through winds and outflows \cite[]{sca05,cat05}. To understand how galaxies form, we therefore need to understand how supermassive BHs and galaxies co-evolve. 

The most important observational tools for studying the co-evolution of BHs and galaxies are the empirical scaling relations between BH mass and global galaxy properties, such as the \msig\ relation \cite[]{fm2000,geb00a}, a particularly tight relation between BH mass, \mbh, and the central, stellar velocity dispersion, $\sigma_*$. The \msig\ relation suggests a causal connection between the BH and the bulge \citep[although see e.g.,][for a counter argument]{jahnke2011}, and is believed to hold clues to understanding the galaxy formation process. By studying how the BH mass relates to the mass of the bulge (as measured by $\sigma_*$), the galaxy evolution process can be examined across different redshift ranges and for different galaxy types. 
  
Active galactic nuclei (AGN) have become instrumental in this work because they contain actively accreting black holes. Not only are they direct probes of the BH-galaxy co-evolution, but galaxies with AGN are the only galaxies in which we can measure black hole masses in the non-local universe. The method is referred to as the virial method, where the width of broad AGN emission lines such as H$\alpha$ or H$\beta$ is used to measure \mbh\ \cite[e.g.,][]{vestergaard2006}.
Whereas measuring virial BH masses in AGN through broad emission lines is 
relatively straightforward, measuring the host galaxy velocity dispersion is 
hampered by the presence of the very bright nuclei which can completely 
overpower the light from the underlying galaxy.  

Attempts have been made to use the width of [OIII] lines as a proxy for the bulge potential \cite[]{nelson2000,shields2003,salviander2007}, but these studies suffer from a large scatter due to non-gravitational influences on the [OIII] line width \cite[such as outflows, radio luminosity, accretion rates, etc; see, e.g.,][]{netzer2007}.  
Studies of low-luminosity AGN at $z\lesssim0.1$ have shown that they follow the \msig\ relation of local, more massive spheroids
 \cite[e.g.,][]{geb00b,nelson2004,greene2006a,woo2010,bennert2011a}, and there is evidence that higher redshift AGN show evolution with regard to the local \msig\ relation.

\cite{woo2006,woo2008} and \cite{treu2007} studied two samples of Seyfert 1 galaxies at $z=0.36$ and $z=0.57$ and found an offset from the local relation
 suggesting that, for a fixed black hole mass, higher redshift spheroids have 
smaller velocity dispersions than local ones. This indicates that, in these 
objects, the galactic bulge is still forming around the black hole. \cite{hiner2012} find a similar offset with respect to the local relation in a sample of six post-starburst AGN at $z\sim0.3$.  These results have serious implications and may provide important observational constraints for galactic evolution models which attempt to explain the co-evolution of black holes and bulges.  The studies by \citeauthor{woo2008}, \citeauthor{treu2007}, and \citeauthor{hiner2012} suggest a very recent growth, within the last 5$-$6 billion years, of intermediate-mass bulges for a given black hole mass, in contrast to predictions made by detailed hydrodynamic simulations \cite[e.g.,][]{robertson2006b}.  
On the other hand, \cite{shen2008} study a large sample of AGN observed with the Sloan Digital Sky Survey (SDSS) at $z<0.452$ and find no evidence of evolution in the relation.   

The study of the \msig\ relation has been severely limited in QSOs due to the observational challenges such a project poses. However, studying \msig\ in QSOs would be desirable since these are potentially the objects where the most rapid growth (in both BH and bulge) occurs.  Studies of \msig\ at $z>0$ need to include QSOs in order to probe the higher end of the relation and, potentially, episodes when either the bulge or the BH is catching up in its growth.  

Measuring velocity dispersions from absorption lines in QSO host galaxies is observationally challenging.  The relatively few studies of host galaxy stellar absorption line spectroscopy in the literature \cite[e.g.,][]{canalizo2001,nolan2001,jahnke2007,wold2010} are not able to recover host spectra in the central regions of the galaxy.  Using these spectra to measure velocity dispersions could lead to ambiguous results since it is difficult to determine whether the dispersions are being measured in the bulge of the galaxy or in the outskirts or some extended tidal feature where \sig\ may not be representative of the mass of the bulge or that of the BH.  Indeed, few attempts have been made to measure \sig\ in the host galaxies of luminous QSOs \citep[e.g.,][]{wolf2008,rothberg2012}, but these studies have been in relation to the host structural properties rather than \msig.
Therefore, it would be ideal to find a class of objects with a natural ``coronagraph'' that would occult the QSO nucleus when measuring \sig, but one that we could remove in order to measure \mbh\ from the width of the broad lines. As described below, we have found that dust-reddened QSOs fit the bill.


Red QSOs have some of the same characteristics as their blue counterparts, such as (at least some of the) strong broad emission lines and high bolometric luminosities, but with much redder continua. At present, there is not a clear definition for red QSOs, so that the different objects that are cataloged as red QSOs do not form a homogeneous class.  The majority of red QSOs discovered to date, however, appear to be reddened by dust, and thus they are considered the dust-obscured equivalent of the blue QSO population \citep[e.g.,][]{cutri2002,marble2003,hall2002,glikman2004,white2003}. Even then, there are cases where the dust is intervening rather than intrinsic to the QSO \cite[e.g.,][]{gregg2002}. Here we focus on those objects where the dust is presumably near the nucleus.

The heavy extinction of the QSO nuclei at optical wavelengths makes dust reddened QSOs excellent candidates to study velocity dispersions and stellar populations in their host galaxies. 
The host galaxy spectrum suffers from much lower (if any) extinction than the nucleus, and stellar absorption features such as those around the Mg\,Ib and G-band regions are clearly visible in the blue region, being almost free of contamination from QSO light.   At longer wavelengths, however, the contribution from the reddened QSO continuum increases and broad H$\alpha$ is clearly visible.   Thus, \sig\ can be measured from features in the underlying host galaxy and \mbh\ can be estimated from the width of H$\alpha$ using the {\it same} spectrum.

In this paper, we present a pilot study to test the feasibility of using these
objects to measure \msig. In \S~\ref{observations} we describe the sample, imaging and spectroscopic observations, as well as the data reduction.  In \S~\ref{analysis} we describe our method to fit stellar velocity dispersions and to measure \mbh.  In \S~\ref{results} we present our results and consider potential biases in our sample, while in \S~\ref{discussion} we discuss our results in the context of other studies of \msig\ that use AGN.
Throughout the paper, we have assumed a cosmological model with $H_{0} = 71$ km\,s$^{-1}$Mpc$^{-1}$, $\Omega_m=0.27$ and $\Omega_{\Lambda} = 0.73$ \citep{spergel2003}, except where noted.  We use the Vega magnitude system unless otherwise specified.

\section{Observations and Data Reduction}\label{observations}

\subsection{The Sample}\label{sample}

Our sample is drawn from the sample of 29 red 2MASS QSOs studied by \citet{marble2003}, which in turn is drawn from the sample compiled by \citet{cutri2002}.  We focused on the subsample from \citet{marble2003} because these objects have Hubble Space Telescope ($HST$) imaging observations that are useful to measure the nuclear luminosities and to study the properties of the host galaxies.
The objects in this sample have $M_{K}\lesssim-25$, $J-K_{s}>2.0$ and detections in each of the three 2MASS bands, $JHK_{s}$.  They have a median redshift of $z=0.213$ and have been spectroscopically confirmed to contain an AGN \citep{smith2002}.  

A fraction of these objects have spectropolarimetry published by \citet{smith2003}.  We avoided the objects that show a blue continuum, with broad emission features dominating the optical spectrum.  Our final sample consists of nine of the remaining targets chosen at random within the observational constraints.

In Table~\ref{table:sample} we list the final sample.  The redshifts listed were measured from stellar absorption lines in our Keck spectra (\S~\ref{spectroscopy}).  $J-K_{s}$ colors and absolute $K_{s}$ magnitudes are quoted from \citet{marble2003}, and the latter assume $H_{0} = 75$ km\,s$^{-1}$Mpc$^{-1}$ and $q_{0}=0.5$.

\begin{deluxetable}{clcccccc}
\tabletypesize{\small}
\tablecolumns{4}
\tablewidth{0pc}
\tablecaption{The Red QSO Sample}
\tablehead{
\colhead {Object} & {QSO} &
\colhead{$z_{host}$} & \colhead{$M_{K_{s}}$} & \colhead{$J-K_{s}$}&\colhead{ESI Exposure} & \multicolumn{2}{c}{Extraction aperture}  \\
\colhead {ID} & {(2MASSi J)} & \colhead{ } & \colhead{ } & \colhead{ } & \colhead{Time (sec)} & \colhead{(arcsec)} & \colhead{(kpc)} \\
}
\startdata
1&005055.7$+$293328 & 0.1356 & $-$25.6 & 2.1 & 1800   & 0.21 & 0.50 \\
2&015721.0$+$171248 & 0.2139 & $-$27.0 & 2.7 & 7200   & 0.21 & 0.75\\
3&022150.6$+$132741 & 0.1398 & $-$25.7 & 2.4 & 5400   & 2.86 & 6.96\\
4&034857.6$+$125547 & 0.2112 & $-$26.7 & 3.3 & 3600   & 0.79 & 2.70\\
5&163736.5$+$254302 & 0.2772 & $-$26.5 & 2.3 & 5400   & 0.21 & 0.88\\
6&165939.7$+$183436 & 0.1709 & $-$26.5 & 2.2 & 5400   & 0.29 & 0.84\\
7&225902.5$+$124646 & 0.1989 & $-$25.6 & 1.9 & 5400   & 0.42 & 1.36\\
8&230442.4$+$270616 & 0.2370 & $-$25.4 & 2.1 & 7200   & 0.97 & 3.61\\
9&232745.6$+$162434 & 0.3664 & $-$27.0 & 2.4 & 5400   & 1.29 & 6.51\\
\enddata
\label{table:sample}
\end{deluxetable}

\subsection{Spectroscopy}\label{spectroscopy}
We obtained deep, medium resolution spectroscopic observations with the Echellette Spectrograph and Imager \cite[ESI;][]{sheinis2002} on the Keck II telescope during three separate nights in October 2003, July 2004 and September 2004.  We observed in the echellette mode, which provides a wavelength coverage from 3900 to 11000 \AA\ in 10 orders (6 through 15).  We used a 1\arcsec\ slit, which projected to $\sim$7 pixels (ranging from $\sim$8 pixels at the short-wavelength end to $\sim$6 pixels at the long-wavelength end) on the MIT-Lincoln Labs 2048$\times$4096 CCD detector.   The spectral resolution is 11.4 km s$^{-1}$ pixel$^{-1}$.
We placed the slit through the center of the host galaxies to ensure that the velocity dispersions that we measure correspond to those of the host galaxy bulges.

All targets were observed under clear weather conditions and subarcsecond seeing ($\sim 0\farcs6$ in V), except for 230442.4$+$270616 and 022150.6$+$132741, which were observed through clouds, and 232745.6 $+$162434, which was observed in high wind conditions.  Exposure times for each target ranged from 1800 to 7200~s; specific times are listed in Table~\ref{table:sample}. 

We also observed a suite of template stars with stellar types ranging from F0 to M6 (listed on Table~\ref{table:templates}) and several spectrophotometric standards.

\begin{deluxetable}{lc}
\tabletypesize{\small}
\tablecolumns{2}
\tablewidth{0pc}
\tablecaption{Stellar templates used for the velocity dispersion fitting}
\tablehead{
\colhead{Star} & \colhead{Luminosity}  \\
\colhead{}     & \colhead{class}
}
\startdata
 HD\,218140   & F0     \\
 SAO\,90936  & F5    \\
 HD\,11851  & F8   \\
 BD$+$264556   & G0   \\
 HD\,10995    & G2IV \\
 SAO\,91026   & G5 V  \\
 SAO\,090989  & K0    \\
 SAO\,91028  & K2    \\
 HD\,11326    & K2III  \\
 HD\,218113    & K5III  \\
 SAO\,90990   & K5 V \\
 Ci20131 & M0    \\
 HD\,11729     & M6   \\
 SAO\,92712 & K0  \\
 SAO\,92718  & G0  \\
\enddata
\label{table:templates}
\end{deluxetable}

The spectra were reduced using a combination of IDL and IRAF scripts written to meet the specific needs of our program.  After bias subtraction and flat fielding, each order was rectified independently in both spatial and wavelength directions using a modified version of the WMKONSPEC package\footnote{http://www2.keck.hawaii.edu/inst/nirspec-old/wmkonspec/index.html}.   A star was observed at multiple positions across the slit in order to trace and rectify the spatial direction of the spectra.  A wavelength solution, obtained from observations of a Cu-Ar lamp, was used to rectify the wavelength direction.  Sky lines were removed by fitting the background in the two-dimensional spectra.   The spectra were calibrated using spectrophotometric standards from \citet{massey1988} observed with the slit at the parallactic angle.  Each target had two or three individual exposures;  we averaged the spatially corrected spectra using the IRAF task {\it scombine}.   

\begin{figure}[h]
\figurenum{1a}
\begin{center}
\epsscale{1.0}
\plotone{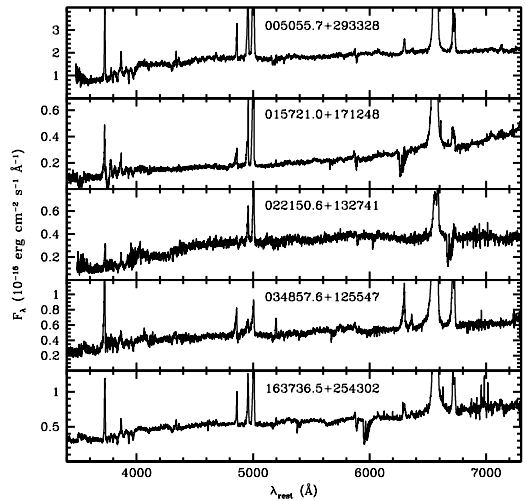}
\caption{Keck ESI rest-frame spectra of our sample of nine 2MASS QSOs.}
\label{figure:spectra1}
\end{center}
\end{figure}

\begin{figure}[h]
\figurenum{1b}
\begin{center}
\epsscale{1.0}
\plotone{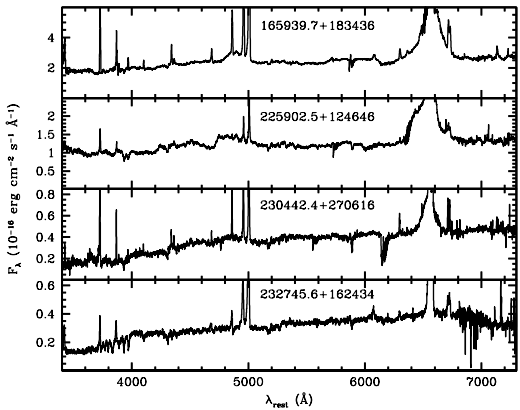}
\caption{Keck ESI rest-frame spectra of our sample of nine 2MASS QSOs.}
\label{figure:spectra2}
\end{center}
\end{figure}

We extracted spectra from the central few kpc of each spectrum, adjusting the size of the aperture to match approximately the effective radius, $r_{\rm eff}$, of each target (see \S~\ref{morphologies}), as listed in Table~\ref{table:sample}.  This corresponded to a slightly different aperture size for each echellette order, since the spatial scale varies for different orders.  We then combined the spectra from all orders into a single spectrum for each target.  We corrected the spectra for Galactic extinction, using the values given by \citet{schlegel1998}.  Finally, we transformed the spectra to rest frame using redshifts measured from stellar absorption lines.  The final spectra for the nine targets are shown in Figs.~\ref{figure:spectra1} and \ref{figure:spectra2}.

\subsection{Imaging}\label{imaging}

The imaging data were obtained from the $HST$ data archive (SNAP-9057; PI: D.\ Hines) and were originally published by \citet{marble2003}.  The targets were observed with the PC chip of the WFPC2 in the F814W filter. Each target had two 400~s exposures.  We combined the images with {\it multidrizzle}, using standard procedures.

\section{Analysis}\label{analysis}

\subsection{Velocity Dispersion}\label{sigma}

The velocity dispersion in the host galaxies was measured by fitting broadened stellar templates to the galaxy spectra. A direct fitting algorithm based on that described by \citet{barth02} was used. The algorithm has been extensively tested and utilized for AGN host galaxies by, e.g.,  \citet{gh06} and \citet{wold07}. The model spectrum $M(x)$ which is fitted to the host galaxy is formed by convolving a stellar template $T(x)$ with a line-of-sight velocity broadening function, assumed to be a Gaussian with dispersion $\sigma_{v}$, adding a power-law continuum $C(x)$ and multiplying by a polynomial $P(x)$: 
\begin{equation}
M(x) = \left\{ \left[ T(x) \otimes G(x)\right] + C(x) \right\} \times P(x),
\end{equation}
\noindent
where $T(x)\otimes G(x)$ is the stellar template convolved with the Gaussian, $G(x)$. The $x$ coordinate is defined as $x=\ln \lambda$ so that velocity shifts are linear in $\ln \lambda$. 
The continuum is $C(x)=c_{0}+c_{1}x$ (note that this is a power-law as a function of wavelength), and $P(x)$ was chosen to be a 3rd degree polynomial 
$P(x)=c_{2}p_{0}(x)+c_{3}p_{1}(x)+c_{4}p_{2}(x)+c_{5}p_{3}(x)$, where $p_{n}(x)$ are Legendre polynomials. In order to find the best-fitting model, we varied the velocity dispersion and the six coefficients $c_{0}$ to $c_{5}$ using a downhill simplex method \citep{ps88} until a global minimum of the $\chi^{2}$ function was found. 

Our object spectra are combinations of QSO and galaxy features, and the featureless continuum $C(x)$ describes the underlying AGN continuum.  The polynomial $P(x)$ accounts for reddening between the template and the object, so the factor $C(x) \times P(x)$ which is added to the convolved stellar template describes the underlying reddened QSO spectrum. By comparing $C(x) \times P(x)$ with an unobscured QSO spectrum, we were able to obtain an estimate of the amount of reddening for each QSO. This is described in \S~\ref{L5100}. 

Stellar templates $T(x)$ were formed from individual and various combinations of the 15 different stellar spectra observed with the same setup as the red QSOs (see Table~\ref{table:templates}). 
Average F, G, K and M spectra were made by normalizing to the flux in the 7000$-$7100 {\AA} region and taking the average of all spectra within each of the four classes.  Since the main source of uncertainty in measuring velocity dispersions typically comes from template mismatches \citep{barth02,gh06}, we ran a series of tests to determine the best template to fit for each region and thus minimize systematic uncertainties, as described in detail below.

Based on previous experience and descriptions of similar analyses in the literature \citep{barth02,gh06,wold07} we chose to estimate velocity dispersions from fits to two different spectral regions: (1) a blue region at 3900$-$4600 {\AA} including Ca\,K but excluding the Ca\,H line (the region 3950$-$3985 \AA\, around Ca\,H was excluded because in some cases AGN contamination from H$\epsilon$ in emission complicated the fit), 
and (2) a red region at 5100$-$5550 {\AA} with the \ion{Mg}{1}b absorption line complex $\lambda\lambda$5167, 5173, 5184 masked out (5155$-$5220 {\AA}). It is well-known that the correlation between [Mg/Fe] and velocity dispersion for elliptical galaxies \citep{jorgensen99,kuntschner01} can cause problems with simultaneous fitting of the \ion{Mg}{1}b absorption and the region redward of it \citep{barth02,wold07}. The red region may also be affected by the AGN \ion{Fe}{2} pseudo-continuum at 5050$-$5520 {\AA}, affecting the fits to this region. This is seen in a couple of our objects, as well as emission from [\ion{N}{1}] $\lambda\lambda$ 5198, 5200. We therefore chose to mask out the region 5155$-$5220 {\AA} from our fits. Apart from this, the chosen red and blue regions are the two most reliable regions in the optical with strong enough stellar features to be used for velocity dispersion measurements. 

The region redward of the \ion{Mg}{1}b contains several iron absorption line features that are well fitted with our composite template. We also obtain good fits to the blue region, in particular to the Ca\,K line at 3933 \AA, H$\delta$-absorption at 4100 \AA, and the G-band at 4300 {\AA}. Generally, we do not see much AGN contamination in this region, except in one case (165939$+$183436), which consequently has larger uncertainties in the velocity dispersion. Some narrow AGN emission lines, such as 
H$\gamma$, H$\delta$, and [\ion{O}{3}] $\lambda$4363, had to be masked from some of the fits, as well as corrupt spectral regions in a couple of cases.  In cases where we could not achieve a good fit to the 4000\AA\ break, but there were enough features longward of the break to constrain the fit, we restricted our fit to a region from 4000$-$4600 \AA .

The galaxy spectra were shifted to the rest frame of the stellar templates, and the fitting algorithm initiated with a reasonable set of parameters (we fit seven free parameters including the velocity dispersion). The parameters were varied freely using the downhill simplex routine until a best fit was found. To improve on the best fit and to ensure that a global minimum was found, the fitting routine was started again, with initial model parameters equal to the first solution but perturbed by 10$-$20 \% in random directions for each parameter. The confidence interval on $\sigma_{v}$ was thereafter found by varying the velocity dispersion in steps of 10 km~s$^{-1}$ while letting the other six parameters float freely. 

In order to minimize systematic uncertainties due to template mismatches, we used single stars as well as different combinations as stellar templates.  For the red region, we find that, in general, we obtain the lowest reduced $\chi^{2}$ when using late stellar types as templates.  Five of the objects were best fit in this region with a K\,5 template.  This is not surprising since this region is generally dominated by features from later stellar types. This also consistent with the results found by Hiner et al.\ (in preparation), who measure stellar velocity dispersions in local galaxies by using templates constructed from a large suite of stellar types, accurately mimicking stellar populations of different ages \citep[see also][]{hiner2012}.  They find a one-to-one correlation in the velocity dispersions that they measure by using a single K star template versus more complex stellar templates in the \ion{Mg}{1}b region (equivalent to our ``red region'').  Only three of the objects in our sample (005055.7+293328, 034857.6+125547, 225902.5+124646) are best fit by including a contribution from earlier stellar types.  Although these fits produced the smallest reduced $\chi^{2}$, the resulting \sig\ is within a few tens of km~s$^{-1}$ of that measured with a K star template.  In fact, the standard deviation in the values of \sig\  measured from different templates is typically smaller than $\sim$20 km~s$^{-1}$.  Since we do not do a detailed analysis of the stellar populations in this work, we do not favor any particular composite template among those that yield the lowest reduced $\chi^{2}$ values.   Therefore, for consistency, we chose to use the same composite template for these three objects as the one we used for the blue region, as described below.   The remaining object, 015721.0$+$171248, shows significant QSO contamination in the red region and we were not able to obtain a reliable fit in this region.

\begin{figure}[hbt]
\begin{center}
\figurenum{2}
\plotone{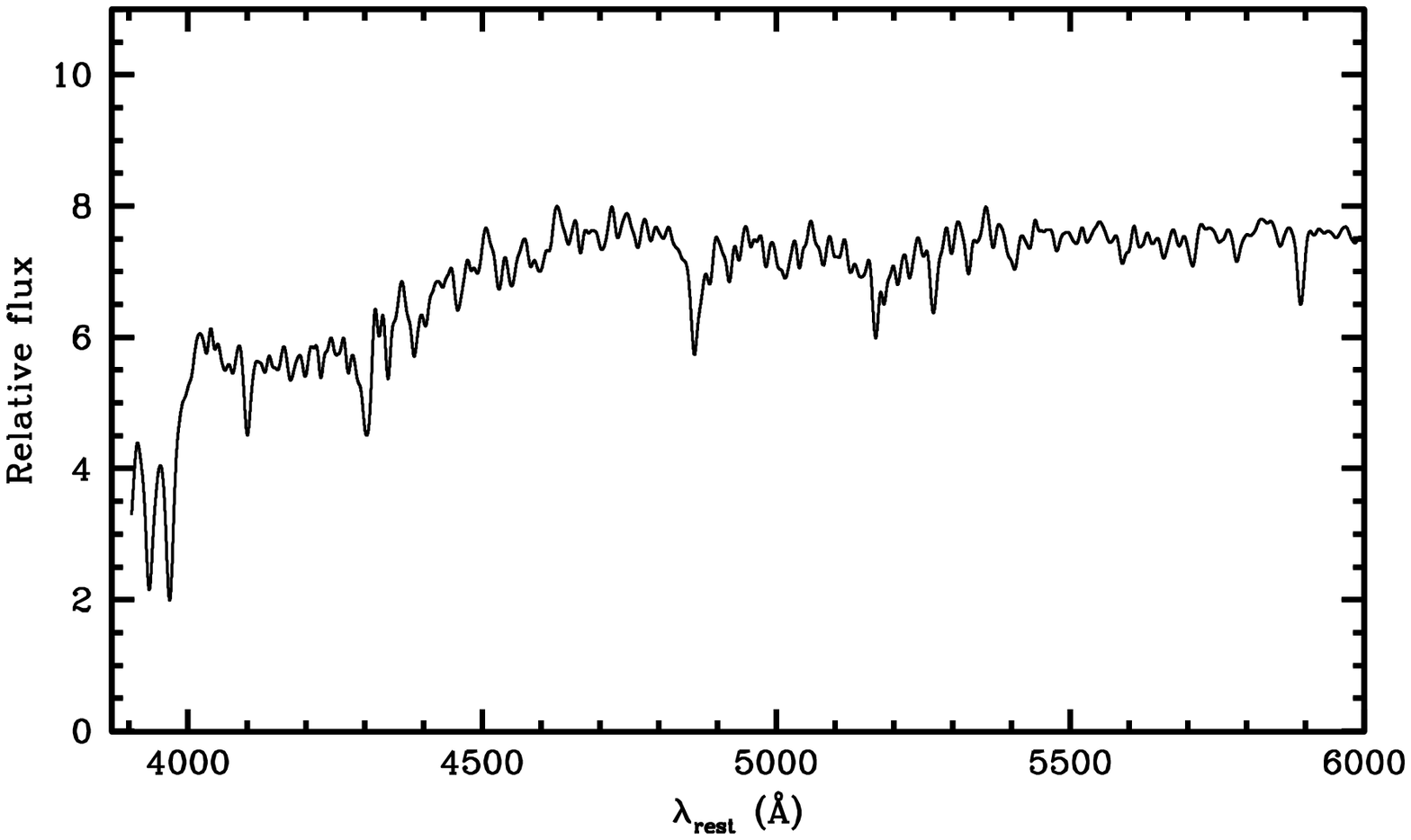}
\caption{Composite stellar template, consisting of 40\% K, 40\% G, 15\% F and 5\% M-type stars (by flux).  The stellar template shown here has been convolved with a $\sigma$ = 165 km s$^{-1}$, which is a typical velocity dispersion of the host galaxies in the sample.}
\label{figure:master_template}
\end{center}
\end{figure}

The spectra of our objects in the blue region show features from both early- and late-type stars.   Therefore, fits using single stellar types as templates consistently resulted in significantly larger reduced $\chi^{2}$ values than those obtained from fits using composite stellar templates.  This is also consistent with the results by Hiner et al., who find that using single stellar types as templates for this spectral region result in systematic uncertainties of several tens of km~s$^{-1}$.  Thus, we tested a few templates consisting of combinations of stellar types with ratios by flux that would mimic intermediate-age populations (i.e., including features from older and younger populations) and chose the one that yield the lowest values of reduced $\chi^{2}$.   Although, in principle, we could do a $\chi^{2}$ minimization to determine the best combination of stellar types for each template, we chose use the same template for every host galaxy in order to reduce the number of free parameters (and degeneracies) in the fits.   The flux contribution at 7000 \AA\ of each stellar type for this composite template is 40\% K, 40\% G, 15\% F and 5\% M-star (Fig.~\ref{figure:master_template}).

\begin{figure}[tb]
\figurenum{3a}
\begin{center}
\includegraphics[scale=0.9]{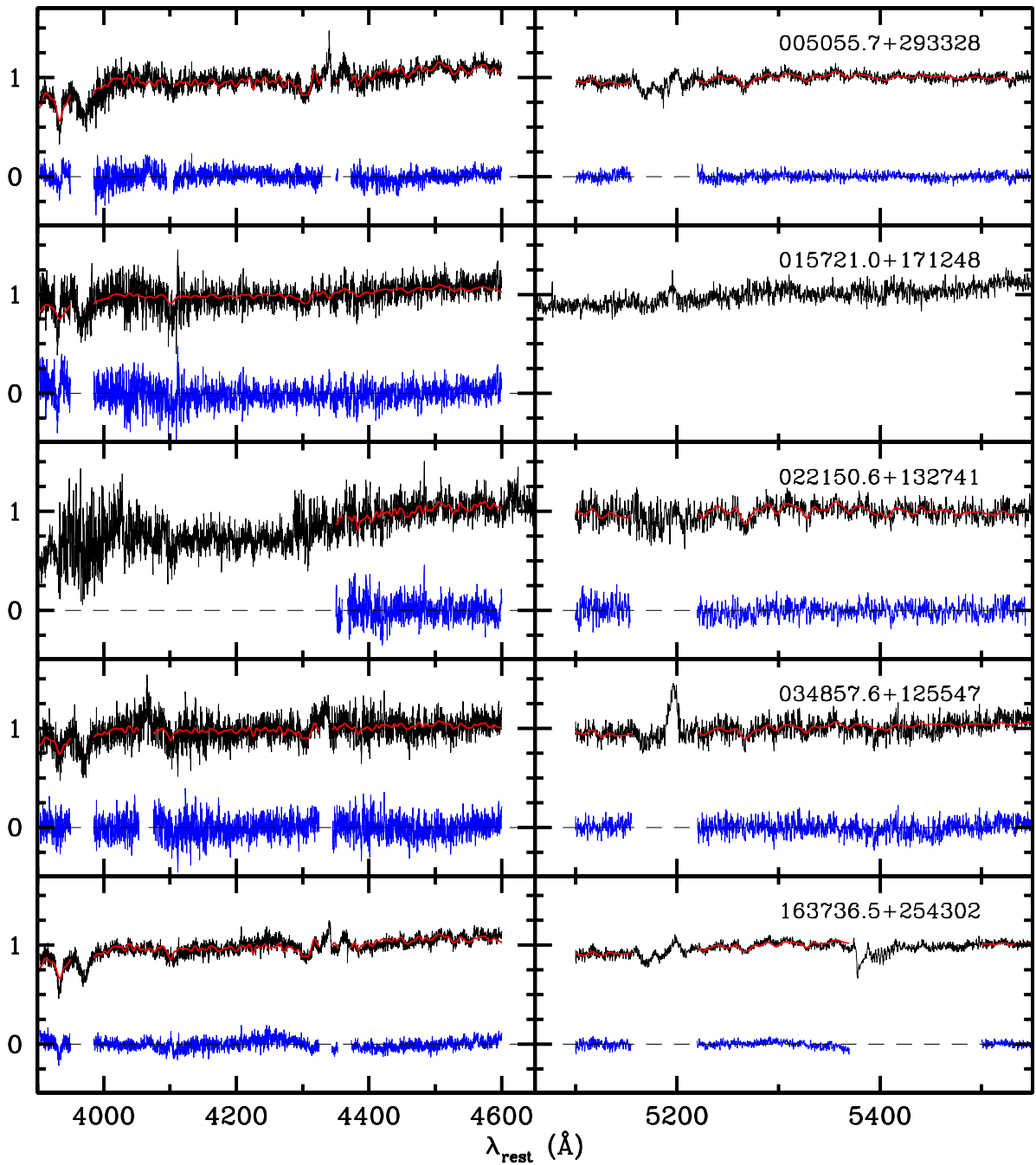}
\end{center}
\caption{Best-fit models (red trace) overplotted on the host galaxy spectra (black trace) in the two spectral regions described in the text. Top of each panel: Rest-frame host spectrum with a fitted broadened stellar template overplotted. Bottom: residual from the fit (blue trace).}
\label{figure:fit1}
\end{figure}

\begin{figure}[tb]
\figurenum{3b}
\begin{center}
\includegraphics[scale=0.9]{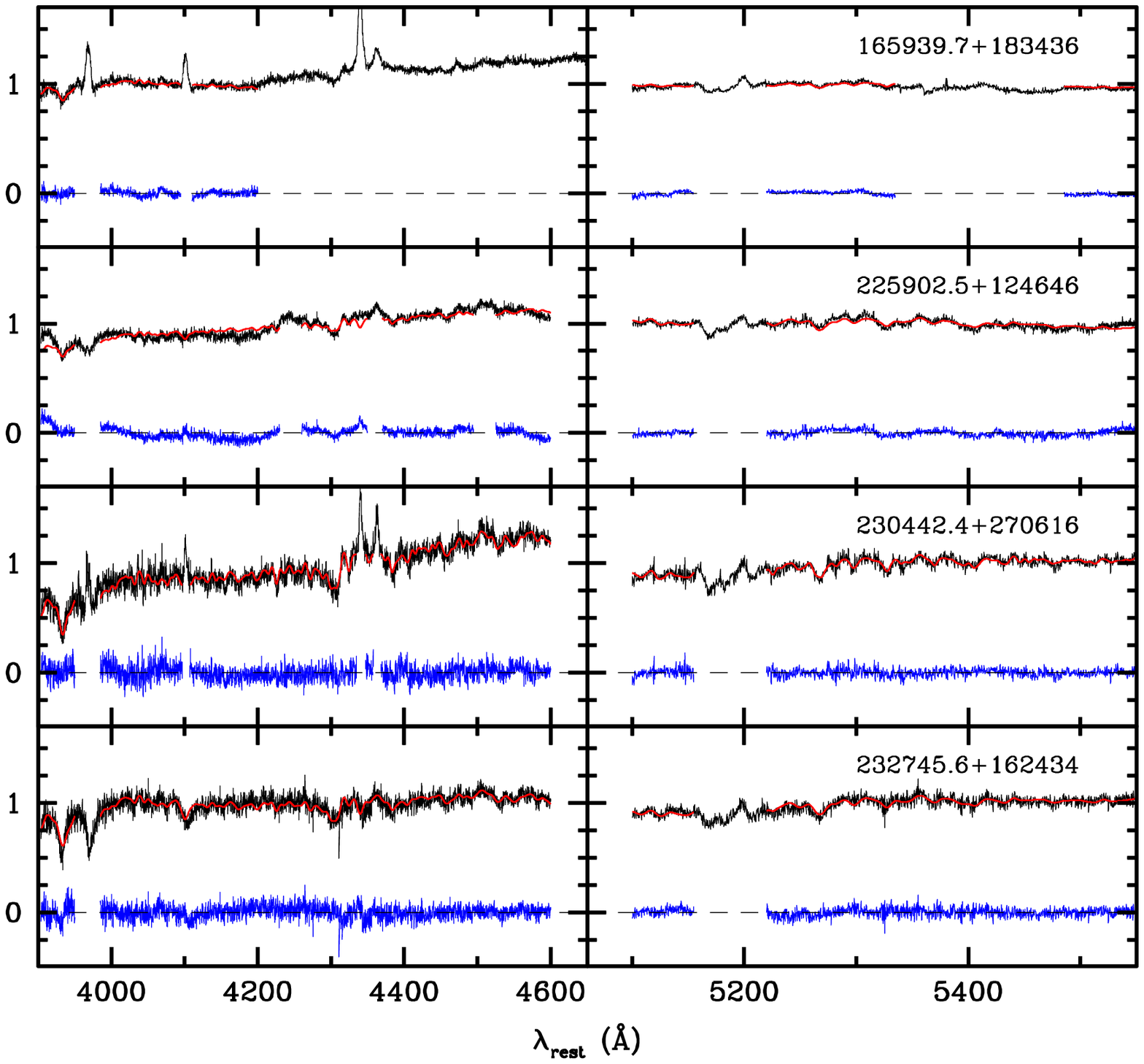}
\end{center}
\caption{Best-fit models (red trace) overplotted on the host galaxy spectra (black trace) in the two spectral regions described in the text. Top of each panel: Rest-frame host spectrum with a fitted broadened stellar template overplotted. Bottom: residual from the fit (blue trace).}
\label{figure:fit2}
\end{figure}

The galaxy spectra in the red and the blue regions with the best-fit models overplotted are shown in Figs.~\ref{figure:fit1} and \ref{figure:fit2}, and the velocity dispersions obtained from these fits are listed in Table~\ref{table:results}.  The reduced chi-squared, $\chi^{2}_{red}$, for each fit is also listed.

The confidence intervals on \sig\ give the uncertainty related to the fitting procedure, with smaller confidence intervals for higher signal-to-noise spectra and for spectra with minimal AGN contribution in the fitting regions. Typically, we obtain 68\% confidence intervals with a width of 10$-$20 km~s$^{-1}$. 
\onecolumn
\twocolumn
\onecolumn
\twocolumn
\begin{deluxetable}{lccccccc}
\tablecolumns{8}
\tablewidth{0pc}
\tablecaption{Best-fit velocity dispersions} 
\tablehead{
\colhead{QSO} & \multicolumn{3}{c}{Blue region} & \multicolumn{3}{c}{Red region} & \colhead{Weighted } \\
\colhead{(2MASSi J)} & \colhead{\sig} & \colhead{$\chi^{2}_{red}$} & \colhead{std. dev.} & \colhead{\sig} & \colhead{$\chi^{2}_{red}$} & \colhead{std. dev.} & \colhead{avg. \sig}
}
\startdata
005055.7$+$293328 &$169_{-4}^{+6\phn}$ & 1.71 & 15.4 & $179_{-15}^{+7}$ & 0.69 & 18.3 & $176_{-17}^{+14}$\\
015721.0$+$171248 &$250_{-8}^{+12}$& 6.75 & 53.0 & \nodata       &\nodata&\nodata& $250_{-54}^{+54}$ \\
022150.6$+$132741 &$115_{-7}^{+9\phn}$ & 1.96 & 40.7 & $140_{-7}^{+7\phn}$  & 3.31 & 20.0 & $129_{-22}^{+22}$ \\
034857.6$+$125547 &$136_{-17}^{+9}$& 6.16 & 26.1 & $157_{-8}^{+9\phn}$  & 2.99 & 17.9 & $152_{-17}^{+17}$ \\
163736.5$+$254302 &$159_{-8}^{+2\phn}$ & 1.49 & 16.7 & $124_{-11}^{+14}$& 0.51 &\phn8.4&$129_{-12}^{+14}$ \\
165939.7$+$183436 &$188_{-22}^{+13}$&0.50 & 22.1 & $150_{-26}^{+32}$& 0.20 & 16.8 & $159_{-25}^{+28}$  \\
225902.5$+$124646 &$198_{-2}^{+10}$& 0.87 & 25.8 & $190_{-12}^{+12}$& 0.69 & 27.8 & $194_{-20}^{+21}$\\
230442.4$+$270616 &$165_{-4}^{+3\phn}$ & 1.53 & 35.3 & $125_{-7}^{+7\phn}$  & 0.76 &\phn3.1& $127_{-7}^{+7\phn}$ \\
232745.6$+$162434 &$229_{-6}^{+7\phn}$ & 2.23 & 21.6 & $195_{-12}^{+11}$& 0.89 & 23.1 & $205_{-19}^{+19}$
\enddata
\tablecomments{Velocity dispersion in units of km\,s$^{-1}$ obtained for the blue and red fitting regions. The errors for the blue and red regions are 68\% confidence intervals; $\chi^{2}_{red}$ is the reduced $\chi^{2}$ for the best fit. The standard deviation in km\,s$^{-1}$ of a series of measurements using different stellar templates is given for each region; this is an estimate of the systematic uncertainty due to template mismatches.   The last column is the average of the \sig\ from the blue and red regions, weighted by their corresponding uncertainties.  The errors given for the average values include the standard deviations added in quadrature.  The blank spaces denote regions where a good fit was not obtainable. Note that the velocity dispersions in this table have not been aperture corrected.}
\label{table:results}
\end{deluxetable}
\noindent
In reality, the uncertainty in the velocity dispersion is larger than this because not all features in the stellar templates used match the host galaxy features equally well.  In order to estimate the uncertainties related to template mismatches we fitted a range of stellar templates to the host galaxies and obtained a standard deviation of the measured values of \sig.   We list the standard deviations for each range of fits for each region in Table~\ref{table:results}.  As mentioned above, the systematic uncertainties due to template mismatches are generally much smaller in the red region than in the blue region, in agreement with the results of Hiner et al.\ (in preparation).    

The velocity dispersions measured from the blue and red regions agree within the uncertainties.  In order to get representative velocity dispersions for each object, we average the two values weighted by their uncertainties and the goodness of fit (as given by $\chi^{2}_{red}$).  The weighted averages with their uncertainties (including systematic uncertainties added in quadrature) are listed in the last column of Table~\ref{table:results}.

\subsection{Galaxy Morphologies}\label{morphologies}
We analyzed the $HST$ imaging data with the two-dimensional modeling program GALFIT \citep{peng2002} to achieve three primary goals: (1) to obtain photometry of the nucleus to be used in the measurement of \mbh, (2) to determine the size and characteristics of the bulge component, if present, and (3) to search for signs of interaction in order to properly interpret our results. 

It is imperative that the appropriate PSF is used when modeling galaxy morphologies. Empirical PSFs are ideal for this purpose, as they reflect the same instrumental conditions that were used for the observations. We searched the $HST$ archive for high signal-to-noise stars (typical peak counts were $\sim2100$) that were observed with the same instrument and filter and that fell near the same CCD position as the QSOs. The input PSF image must be large enough to encompass the wings of the function, so we chose a large region and interpolated across extraneous sources within the region. Since the PSF is undersampled in the PC image, we broadened it with a Gaussian of $\sigma = 0.8$ pixels. This increased the full-width at half-maximum (FWHM) of the PSF to be greater than 2 pixels, while preserving the encompassed flux.

The science images need to be large enough to accurately estimate the background sky level. This is especially important, because the edges of galaxy profiles are sensitive to the background sky level, and can affect, for example, the S\'{e}rsic index, $n$, of the profile fit to the galaxy. We cut each image to 15''.75 by 15''.75 (350 pixels x 0.045 arcsec pix$^{-1}$) centered on the science target, which is several times the size of the targets themselves. This large region sometimes included other objects, which we either masked from the fit or fit with an extra component. To match the resolution of the PSF, we also broadened the science images with the same Gaussian used on the PSF. 

We followed an iterative process while fitting the morphologies of the galaxies. Each fit included the point source (a scaled PSF) and background sky level estimated by GALFIT. We began with simple models of single bulge or disk components for the host galaxies and added complexity as necessary. All but one of the galaxies required two or more components to model the host. Where the image showed multiple objects in the field, the secondary object was fit with its own component or masked from the fit. We generally used free index S\'{e}rsic profiles to model the bulges, but in many cases, GALFIT did not converge on a reasonable S\'{e}rsic index. To fit these hosts, we ran several iterations of GALFIT holding the index constant at values $n =$ 1, 2, 3, and 4. In each case the $n = 4$ fit the image the best. We adopted the best-fit models based on the $\chi^{2}$ values and residuals of the model subtracted image.

We report magnitudes for the best fits of each of the targets. Visual inspection of the residual images indicates that the point source and PSF were slightly mismatched.  In order to estimate the uncertainty in the magnitudes, we also performed a direct PSF subtraction and determined the upper and lower limits for which the PSF was clearly over- and under-subtracted, respectively.  The magnitudes measured by direct subtraction were, in each case, within 0.1 mag of the value obtained from the GALFIT fits, except for 005055.7$+$293328, for which the difference was 0.36 mag.   Our measured magnitudes are systematically fainter than those reported by  \cite{marble2003}, with the differences ranging from less than 0.1 mag to 1.1 mag in the most extreme case (165939.7$+$183436).  The differences likely arise from the fact that we fit simultaneously a PSF with the different components of the host galaxy (e.g., bulge and disk), so the nuclear fluxes that we measure are less likely to be contaminated by light from the host galaxy.  Another reason for the differences in magnitudes may be the choice of PSFs: Whereas we selected empirical PSFs from the $HST$ archive, \citet{marble2003} used one of the QSOs in their sample as the PSF to subtract from all the other QSOs.  They chose 222202.2$+$195231 since it did not show significant extended emission after subtracting an artificial PSF created with the software package Tiny Tim \citep{krist2001}.

In Table~\ref{table:galfit} we list the magnitudes of the PSF and S{\'e}rsic components used to fit each object.  The magnitudes are apparent $HST$ F814W magnitudes, not corrected for Galactic extinction nor K-corrected, to allow for direct comparison with results published by \cite{marble2003}.  Every object appears to have a clear bulge component.  In Figs.~\ref{figure:galfit}, \ref{figure:galfit_b} and \ref{figure:galfit_c} we show the image, the best fit model, and the residuals after subtraction of each of the targets.   One-dimensional surface brightness profiles of each object, including all the components used to model them, are shown in Fig.~\ref{figure:radial}.
\begin{deluxetable}{llcccc}
\tablecolumns{12}
\tablewidth{0pc}
\tablecaption{Results of Modeling the QSO Host Galaxies Using GALFIT}
\tablehead{
\colhead{QSO} & \colhead{Component} & \colhead{$m_{\rm F814W}$} & \colhead{$r_{\rm eff}$} & \colhead{$r_{\rm eff}$} & \colhead{S{\'e}rsic Index}\\
\colhead{(2MASSi J)} & & \colhead{(mag)$^{a}$} & \colhead{($\arcsec$)} & \colhead{(kpc)} & \\
\colhead{(1)} & \colhead{(2)} & \colhead{(3)}  & \colhead{(4)} & \colhead{(5)}
& \colhead{(6)}
}
\startdata
005055.7$+$293328 & PSF   & 20.50 & \nodata & \nodata & \nodata \\
                  & bulge & 19.27 & 0.22 & 0.52 & 4 (fixed)     \\
                  & disk? & 19.42 & 3.46 & 8.22 & 0.15 \\
                  & arms  & 18.56 & 1.89 & 4.50 & 0.63 \\
\hline
015721.0$+$171248 & PSF   & 20.45 & \nodata & \nodata & \nodata \\
		  & bulge & 20.67 & 0.19 & 0.67  & 4 (fixed) \\
                  & disk  & 19.86 & 0.72 & 2.50  & 1 (fixed) \\
\hline
022150.6$+$132741 & PSF   & 21.17 & \nodata & \nodata & \nodata \\
		  & bulge & 17.93 & 2.89 & 7.04 & 4 (fixed) \\
		  & disk  & 18.88 & 2.13 & 5.19 & 0.5 \\
\hline
034857.6$+$125547 & PSF   & 21.57 & \nodata & \nodata & \nodata \\
		  & bulge & 19.80 & 0.79 & 2.70 & 5.8 \\
		  & companion  & 20.88 & 0.44 & 1.51 & 1 (fixed) \\
		  & companion  & 20.08 & 0.70 & 2.40 & 1 (fixed) \\
\hline
163736.5$+$254302 & PSF   & 21.37 & \nodata & \nodata & \nodata \\
		  & bulge & 19.88 & 0.22 & 0.92 & 4.7 \\
		  & disk  & 19.87 & 0.49 & 2.04 & 0.6 \\
\hline
165939.7$+$183436 & PSF   & 19.11 & \nodata & \nodata & \nodata \\
		  & bulge & 18.21 & 0.26 & 0.76 & 4 (fixed) \\
\hline
225902.5$+$124646 & PSF   & 19.40 & \nodata & \nodata & \nodata \\
		  & bulge & 18.82 & 0.41 & 1.33 & 3.7 \\
		  & disk  & 19.51 & 1.20 & 3.89 & 1 (fixed) \\
\hline
230442.4$+$270616 & PSF   & 20.70 & \nodata & \nodata & \nodata \\
		  & bulge & 19.23 & 0.98 & 3.65 & 4 (fixed) \\
		  & disk  & 21.50 & 0.32 & 1.18 & 1 (fixed) \\
\hline
232745.6$+$162434 & PSF   & 22.12 & \nodata & \nodata & \nodata \\
		  & bulge & 19.20 & 1.30 & 6.56 & 4 (fixed) \\
		  & disk  & 20.07 & 1.03 & 5.22 & 1 (fixed) \\
\enddata
\tablecomments{$^{a}$ F814W magnitudes not corrected for Galactic reddening or for intrinsic reddening.}
\label{table:galfit}
\end{deluxetable}

The object 005055.7$+$293328 is hosted by a spiral galaxy and has a prominent dust lane in the central region.  In order to obtain the best estimate for the nuclear flux we fit the host galaxy with multiple components so as to minimize the residuals after subtraction. The best fit included the spiral arms, disk, bulge, and an additional narrow S{\'e}rsic profile (not included in Table~\ref{table:galfit}) to account for the asymmetric light distribution due to the dust lane.  While fitting all these components allows us to obtain a good fit for the PSF, it is likely that we are underestimating the flux and the effective radius of the bulge, since we are splitting it into two components.

For 163736.5+254302 and 165939.7+183436, we use additional components to fit the apparent companions.  These components are not listed in Table~\ref{table:galfit}, but they are shown in Fig.~\ref{figure:radial}.

\vspace{0.6in}

\subsection{Black hole mass estimates}

We estimated virial masses for the BHs in the sample 
by using the scaling relation for \ha\ given by \cite{greene2010}:
\begin{eqnarray}
\mbhg = (9.7 \pm 0.5) \times 10^6 
\left(\frac{\lf}{10^{44}~{\rm erg~s^{-1}}} \right)^{0.519 \pm 0.07} \nonumber \\
 \times \left(\frac{\fwha}{10^3~{\rm km s^{-1}}} \right)^{2.06 \pm 0.06}
\end{eqnarray}

\citeauthor{greene2010} use the virial coefficient measured by \citet{onken2004}, i.e., $<f>$ = 5.5, which is a factor of 1.8 higher than the value assumed for a spherically symmetric broad line region \cite[e.g.,][]{kaspi2000}.  This value is also consistent with that found by \citet{woo2010} from a sample of reverberation mapped AGN.  Although \citet{graham2011} find a lower virial factor (i.e., 2.8), we adopt $<f>$ = 5.5 since this allows us to compare our results directly with those of previous studies (see \S\ref{results}).

In the scaling relation for virial masses, the size of the broad line
region is a function of the FWHM of the broad component of H$\alpha$, FWHM$_{H\alpha}$, and the AGN continuum luminosity at rest frame 5100 \AA, 
$L_{5100}$.  Here we describe our procedure to measure each of these two quantities.

\subsubsection{Measuring FWHM$_{H\alpha}$}

In order to measure the width of the broad \ha\ line, we first had to do a detailed decomposition of the narrow \ha\ component and the narrow \ion{N}{2} $\lambda\lambda$~6548, 6583 lines.  We performed all the line fits using the {\it specfit} task \citep{Kriss94} in the IRAF STSDAS package, and following the method outlined by \citet{glikman2007} and \citet{GH04}, as described below.

First, we created a narrow line model by fitting the \ion{S}{2} $\lambda\lambda$6716,  6731 doublet. The line profiles often showed blue wings.  Thus, we fit each line using two Gaussians, one of which was allowed to have skew (or asymmetry), although we found that skew was not needed in all cases. The continuum was fit simultaneously with a power law function. To ensure the same model fit for both lines, and to avoid unrealistic degeneracies, we constrained the model for each line of the doublet to be the same. This was done by matching the widths of each
\onecolumn
\begin{figure}[tbh]
\begin{center}
\figurenum{4a}
\epsscale{0.9}
\plotone{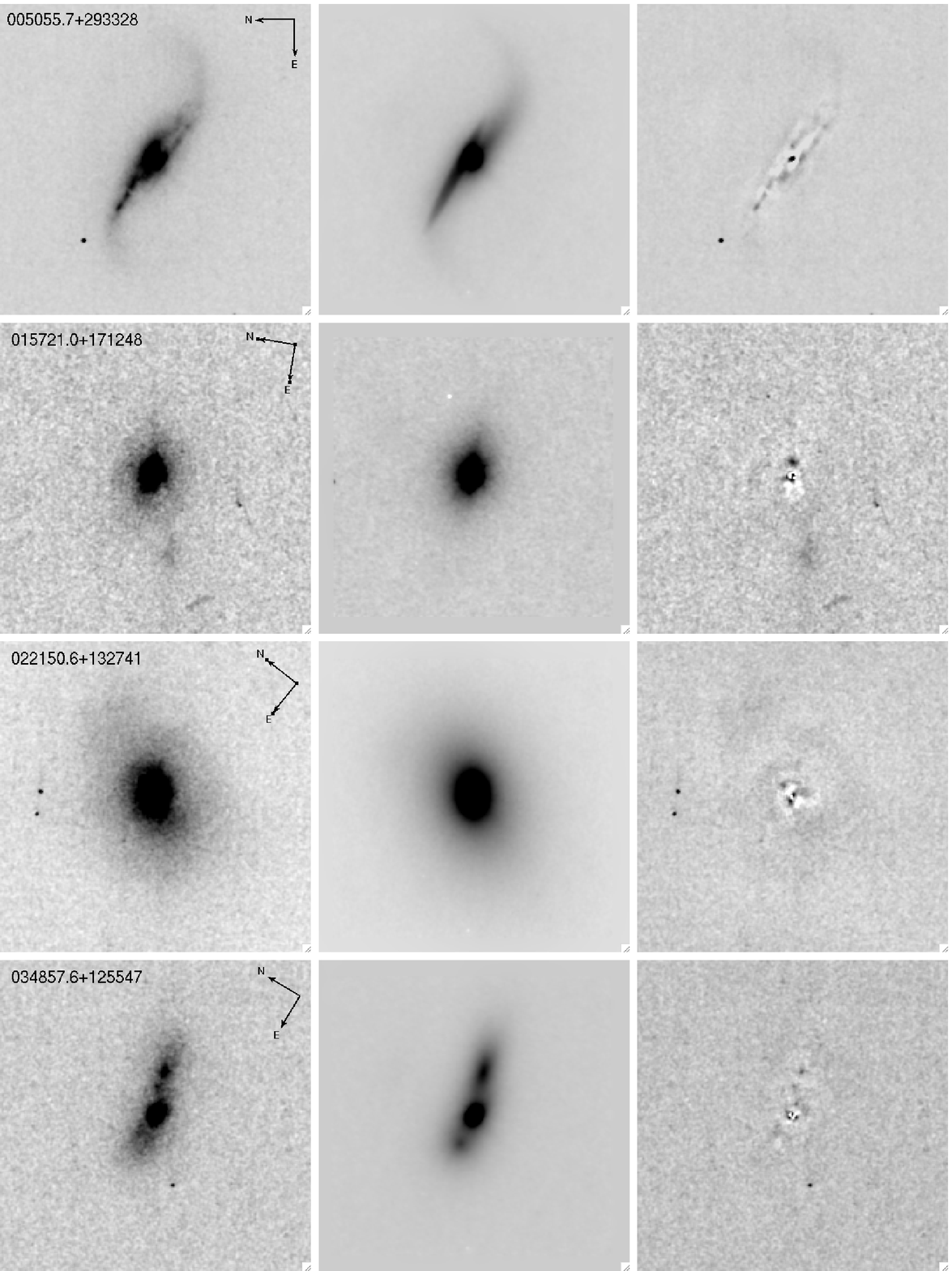}
\caption{$HST$ F814W images of the 2MASS QSOs. For each row, the left panel is the original image, the central panel is the best model obtained with GALFIT, and the right panel is the residuals after subtracting the best model from the observed image.  Each image is 10\arcsec$\times$10\arcsec.}
\label{figure:galfit}
\end{center}
\end{figure}
\twocolumn
\onecolumn
\begin{figure}[tbh]
\begin{center}
\figurenum{4b}
\epsscale{0.9}
\plotone{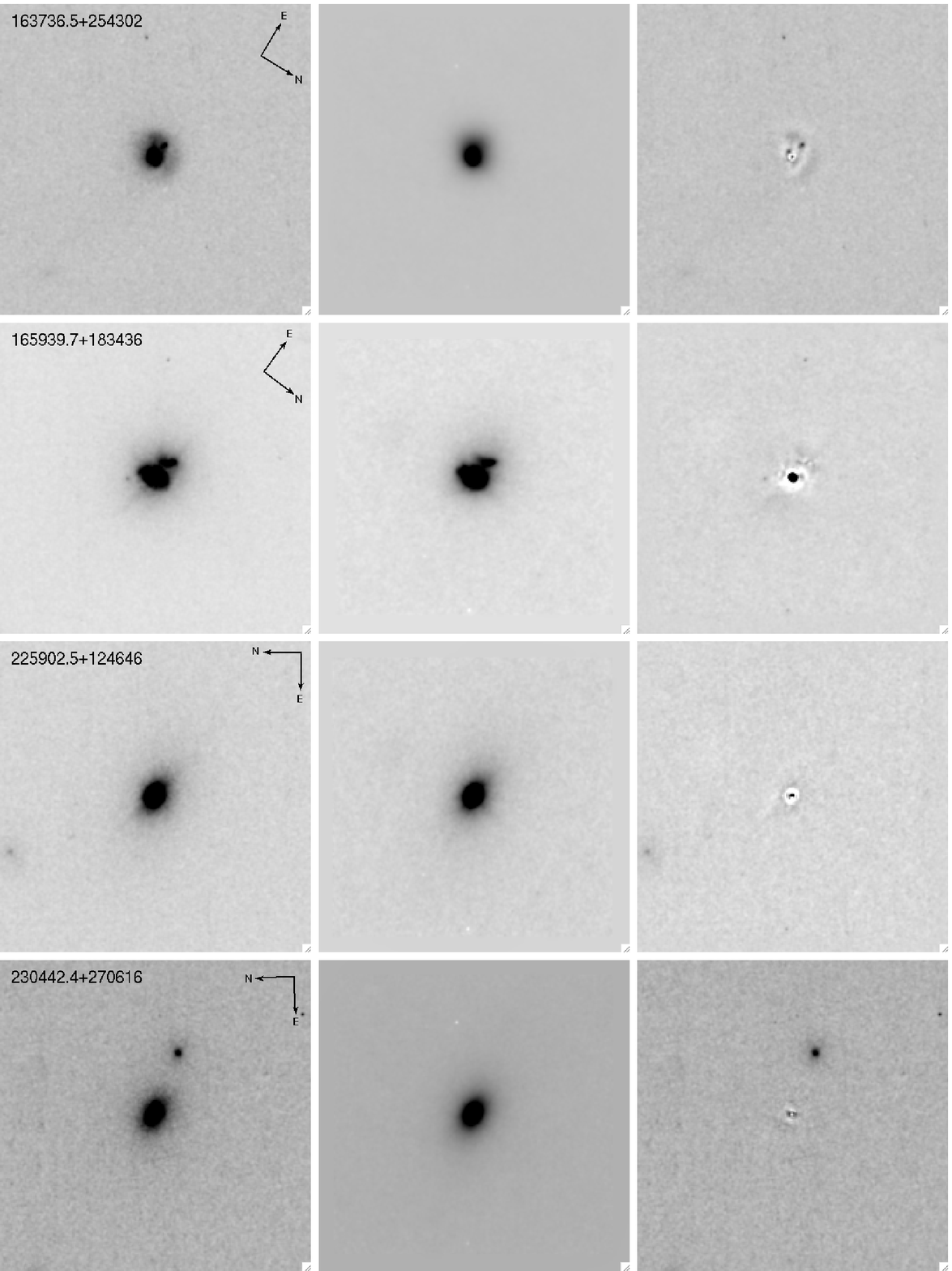}
\caption{Same as Fig.~\ref{figure:galfit}}
\label{figure:galfit_b}
\end{center}
\end{figure}
\twocolumn
\onecolumn
\begin{figure}[tbh]
\begin{center}
\figurenum{4c}
\epsscale{0.9}
\plotone{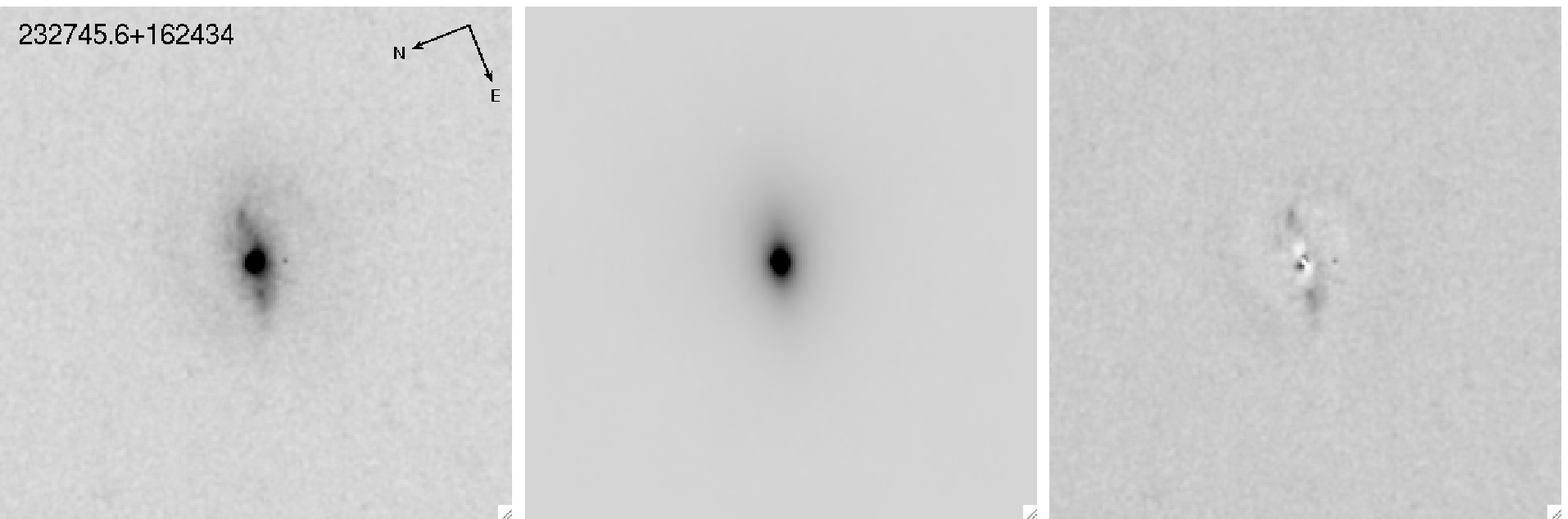}
\caption{Same as Fig.~\ref{figure:galfit}}
\label{figure:galfit_c}
\end{center}
\end{figure}
\twocolumn
\noindent
component and the relative positions of the doublet. Because we used a two-Gaussian model, an extra constraint on the flux ratio was necessary to ensure the same total model for each line. \citet{Garstang52} calculates the transition probability and intensity ratio of the doublet $^{4}$S - $^{2}$D transition in \ion{S}{2}. He finds it can range from 0.5 to 0.8 depending on the parameters used in the calculation. However, he also cites a reasonable observed value of 0.74 based on NGC 7662. We adopted this observed value for our \ion{S}{2} fits, as it produced the most reasonable fit of the range.

In some cases the \ion{S}{2} doublet was contaminated by the \ha\ broad emission or telluric absorption. In these cases we fit the [\ion{O}{3}] $\lambda\lambda$4959, 5007 doublet to create a narrow line model. Similar techniques as with the \ion{S}{2} doublet were used to fit the [\ion{O}{3}] doublet. In this case the doublet flux ratio was constrained to 2.87 \citep{OF2006}, which was a good fit for our data. 

We then fit the region around \ha\ using the narrow line model for \ion{N}{2} $\lambda$6548 and \ion{N}{2} $\lambda$6583. We constrained the relative position of the \ion{N}{2} lines using the laboratory wavelengths for these lines. While \citet{glikman2007} constrained the \ion{N}{2} flux ratio to a theoretical value of 2.96, we found better fits when we let the flux of each \ion{N}{2} line be freely fit. This often resulted in a flux ratio greater than 2.96. In addition, we included a narrow line component of \ha\ using the fitted narrow line model and we used a power law to characterize the continuum. Finally, we included a broad Gaussian (with initial width of 5000 km s$^{-1}$) to model the broad line \ha\ component. We first fixed the broad \ha\ line to be symmetric, then added skew after its centroid position had been located by the fitting routine.  We show the best fits with each of their components in Fig.~\ref{figure:lineprofiles}.

{\it Specfit} uses a simplex method to minimize $\chi^{2}$. We adjusted input parameters and initial step sizes iteratively to probe different local $\chi^{2}$ minima and thus found the global $\chi^{2}$ minimum.
The errors in the FWHM$_{H\alpha}$ resulting from the fit are typically $\sim$15 km s$^{-1}$.  The true uncertainties, however, are much larger since they are dominated by the uncertainties in the modeling of the narrow component and the decomposition of the broad and narrow components.  

In addition, although the AGN is much less extinguished in these wavelength regions than in the regions where we measure \sig, there is still significant contribution from the host galaxy to the continuum around \ha.  We found that subtracting different stellar \ha\ absorption lines (corresponding to different stellar templates or stellar populations) has no measurable effect on the value of FWHM$_{H\alpha}$.  However, the shape of the stellar continuum around \ha\ is different for different stellar populations, and this can potentially affect the fitting of the broad component, especially for the objects with the broadest emission lines.  Thus the choice of stellar template to subtract from the AGN spectrum introduces a systematic uncertainty.  In order to estimate this uncertainty, we measured FWHM$_{H\alpha}$ from the different template-subtracted emission lines and found the standard deviation for the different measurements.   We added this uncertainty in quadrature to the fitting and systematic errors.

The final values for FWHM$_{H\alpha}$ and their uncertainties are listed in Table~\ref{mbh}. The FWHM$_{H\alpha}$ we measure for 225902.5$+$124646 is highly uncertain because H$\alpha$\ has a very asymmetric profile, with very extended emission on the blue side of the line.  This is likely indicative of an outflow, so that the \mbh\ we derive from it may not be reliable.  We attempted a variety of ways to separate the core of the line from the outflow, by fitting the latter with multiple Gaussians blueshifted with respect to the core.  In order to ensure that we obtain a conservative measure of \mbh, we chose the smallest value of FWHM$_{H\alpha}$ that would fit the line.   The upper limit on the errors is given by the FWHM that would be obtained if a single Gaussian were used to fit the broad profile.

We note that some studies, including those of reverberation mapped AGN \citep[e.g.,][]{woo2010} use line dispersions rather than FWHM to calculate \mbh.   However, these measurements are based on $H\beta$ rather than \ha.  For our study, we are unable to measure the line dispersion or FWHM of H$\beta$, since the AGN is still highly obscured at those wavelengths.  Instead, we must use \ha\ as a proxy for H$\beta$.   The relations between these two lines in the literature are currently much better constrained for FWHM than for line dispersions.  For example, the relation between FWHM$_{H\alpha}$ and FWHM$_{H\beta}$ used by \citet{greene2010} is derived by fitting 162 objects 
\onecolumn
\begin{figure}
\begin{center}
\figurenum{5}
\includegraphics[scale=0.27]{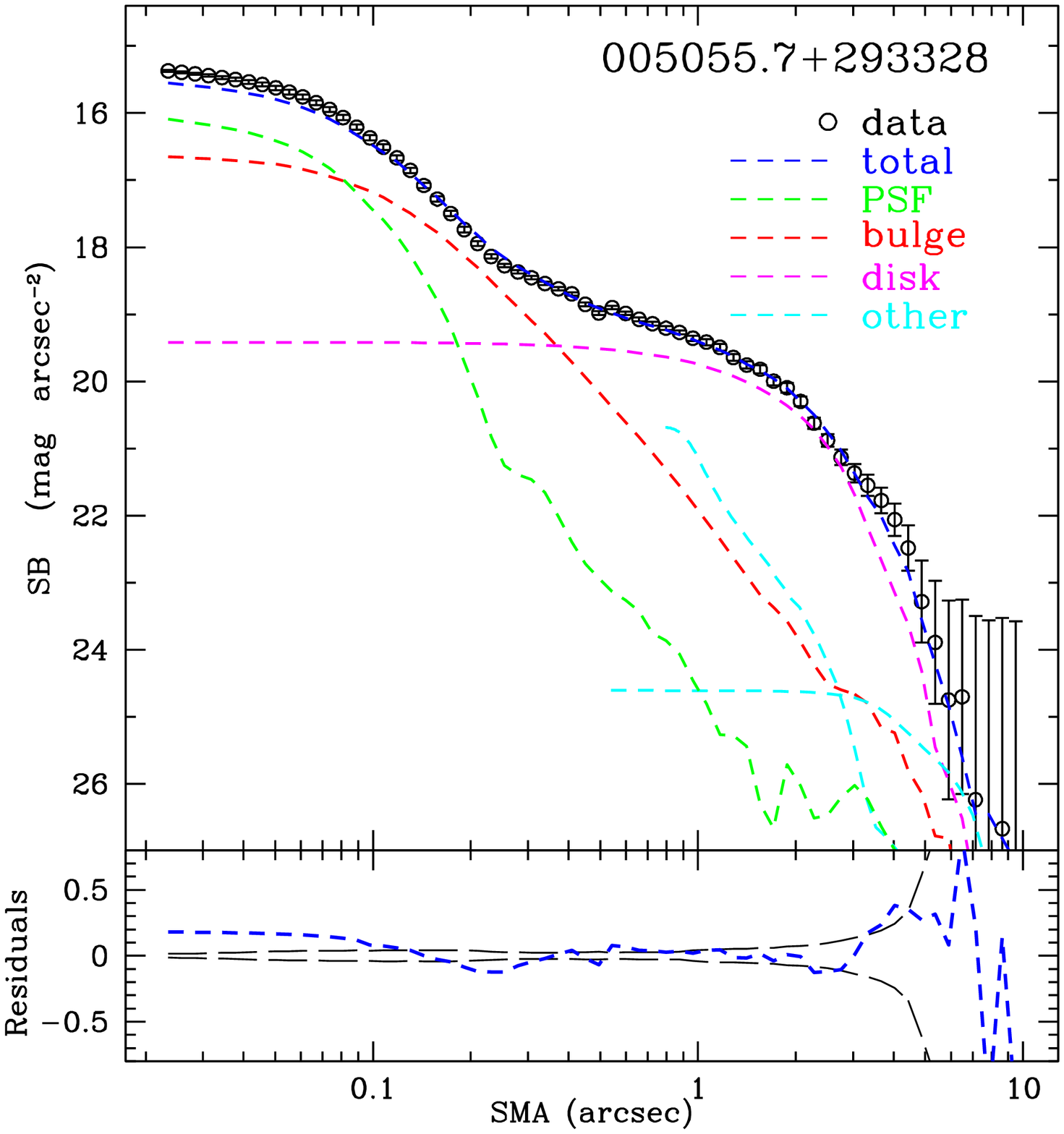}\hspace*{0.1cm}
\includegraphics[scale=0.27]{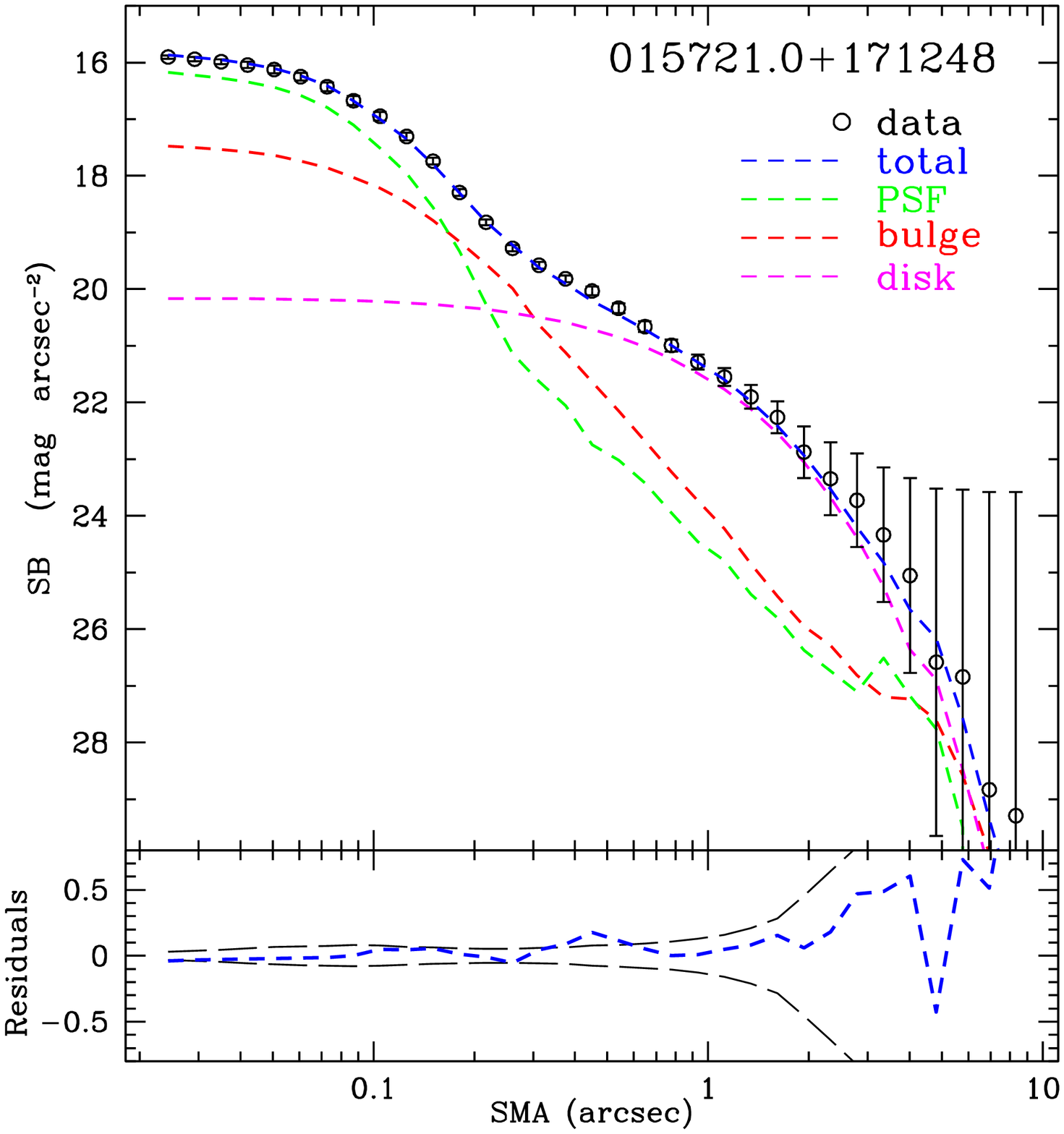}\hspace*{0.1cm}
\includegraphics[scale=0.27]{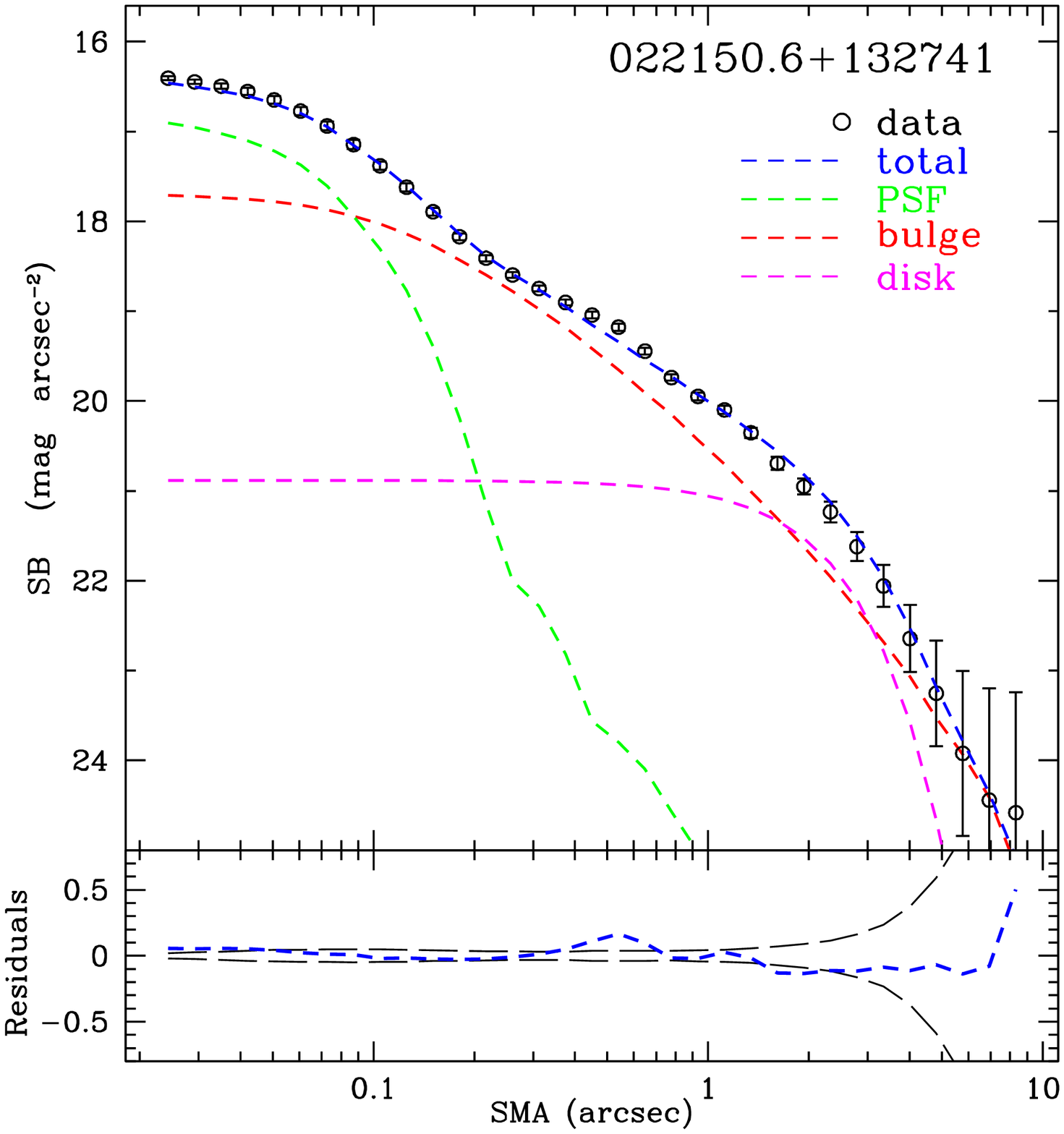}\\
\includegraphics[scale=0.27]{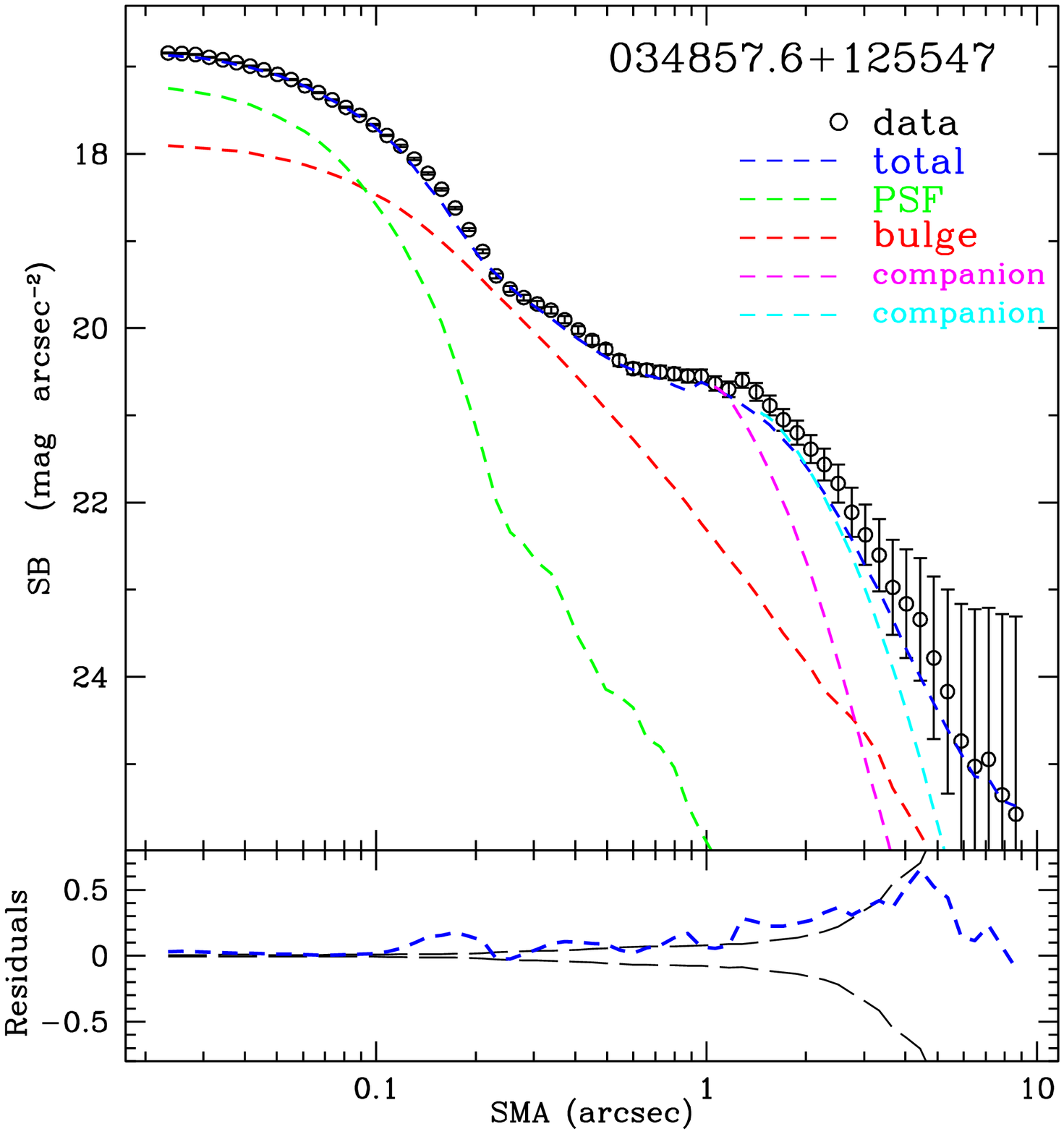}\hspace*{0.1cm}
\includegraphics[scale=0.27]{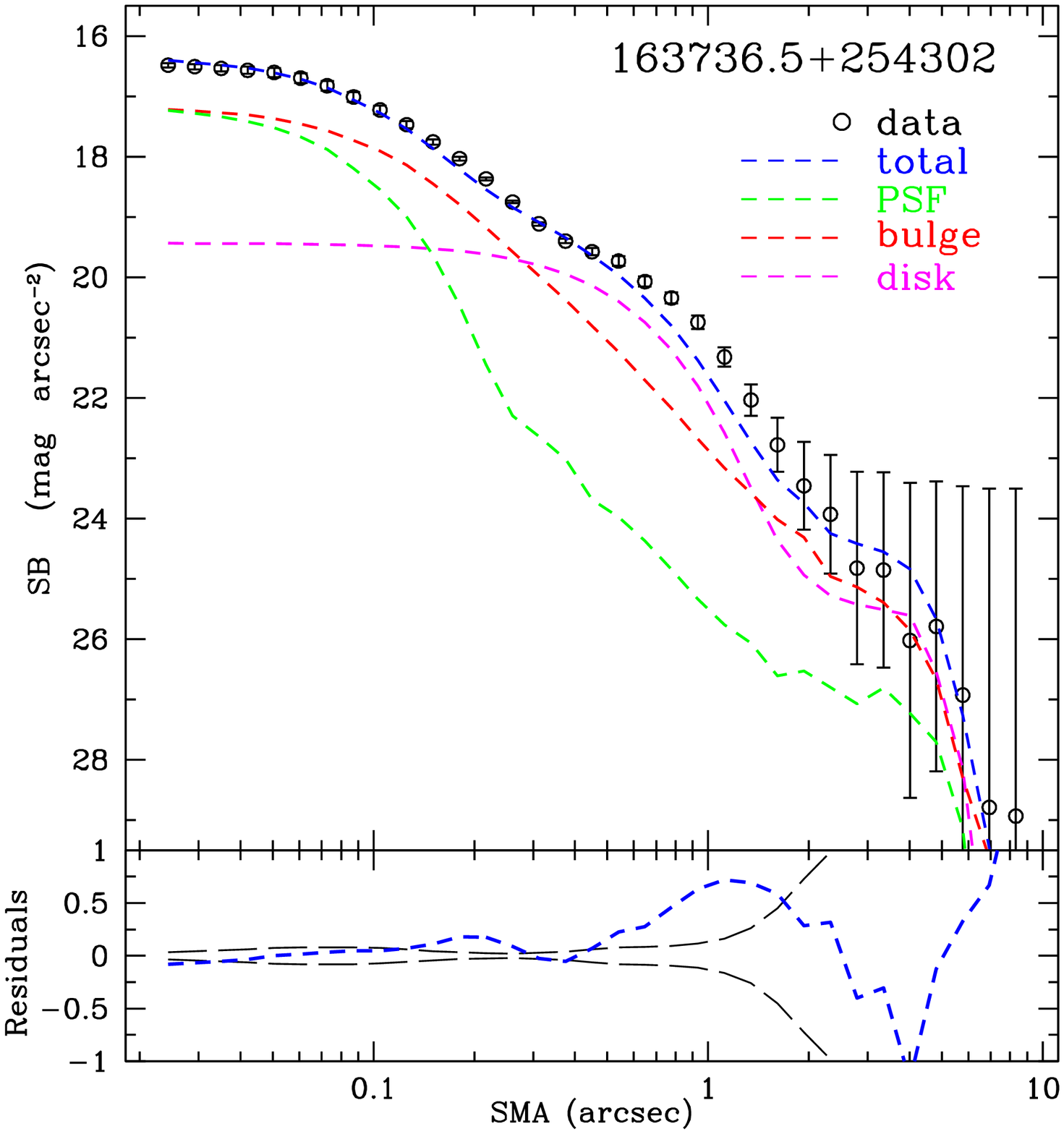}\hspace*{0.1cm}
\includegraphics[scale=0.27]{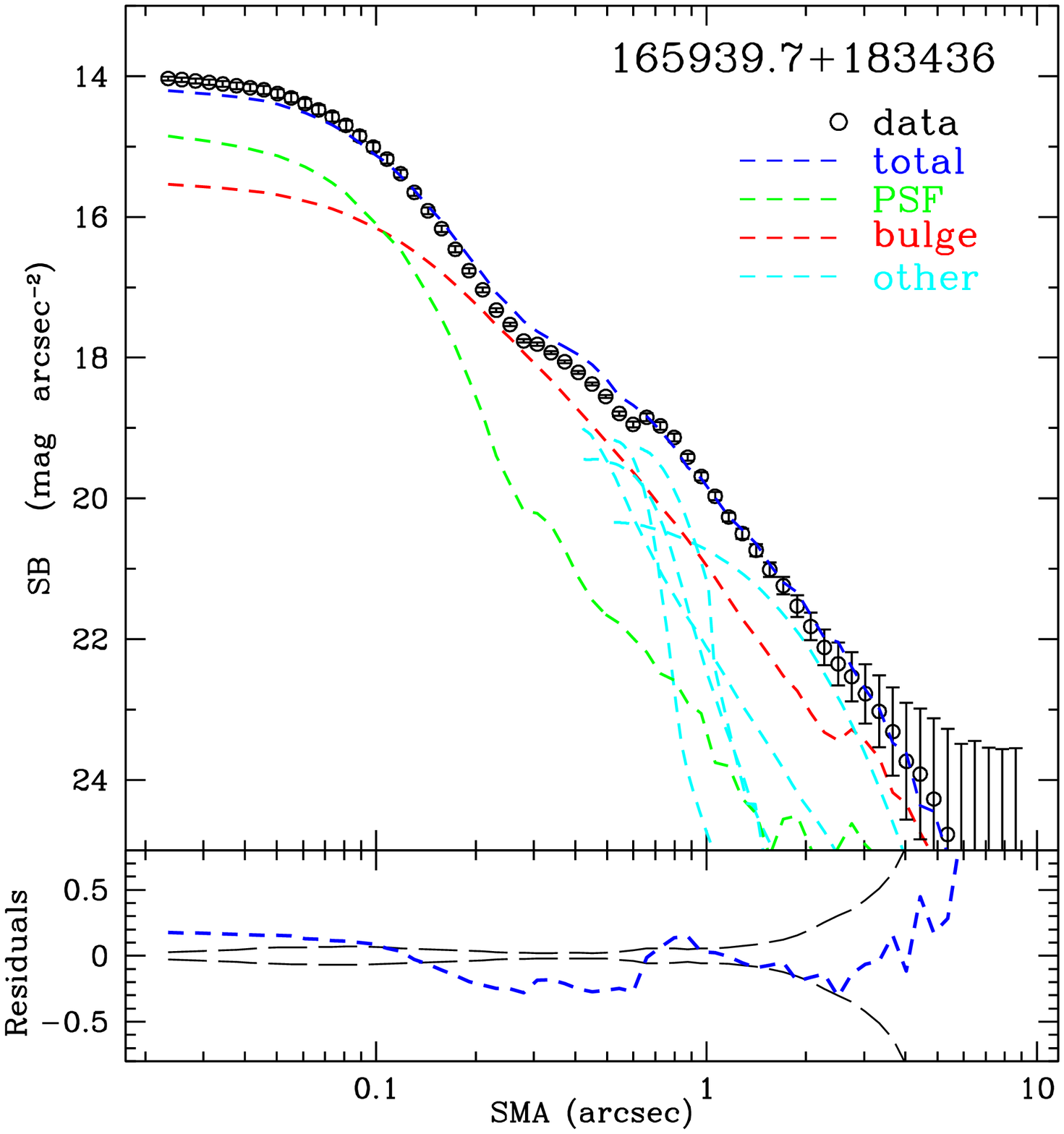}\\
\includegraphics[scale=0.27]{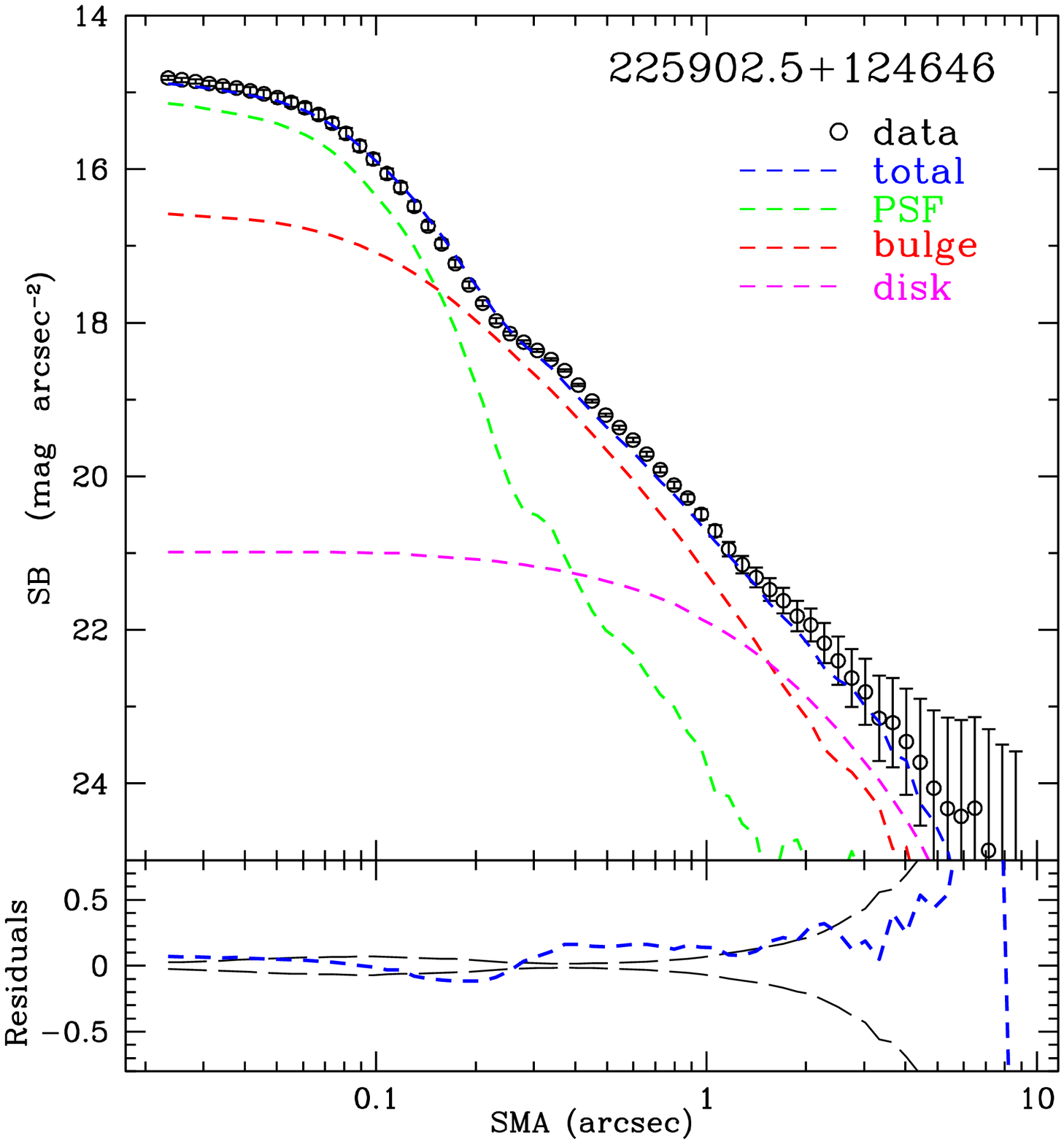}\hspace*{0.1cm}
\includegraphics[scale=0.27]{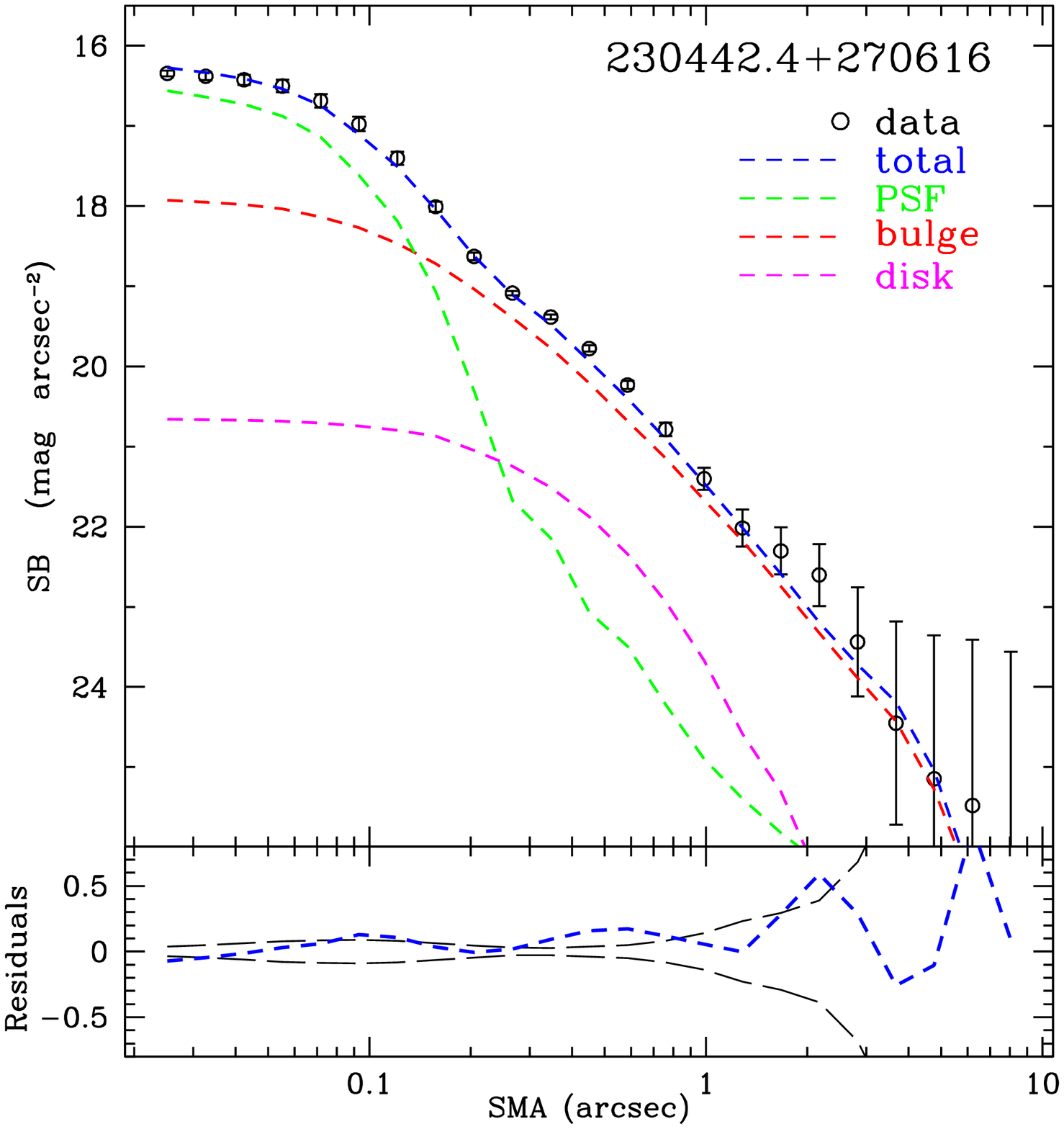}\hspace*{0.1cm}
\includegraphics[scale=0.27]{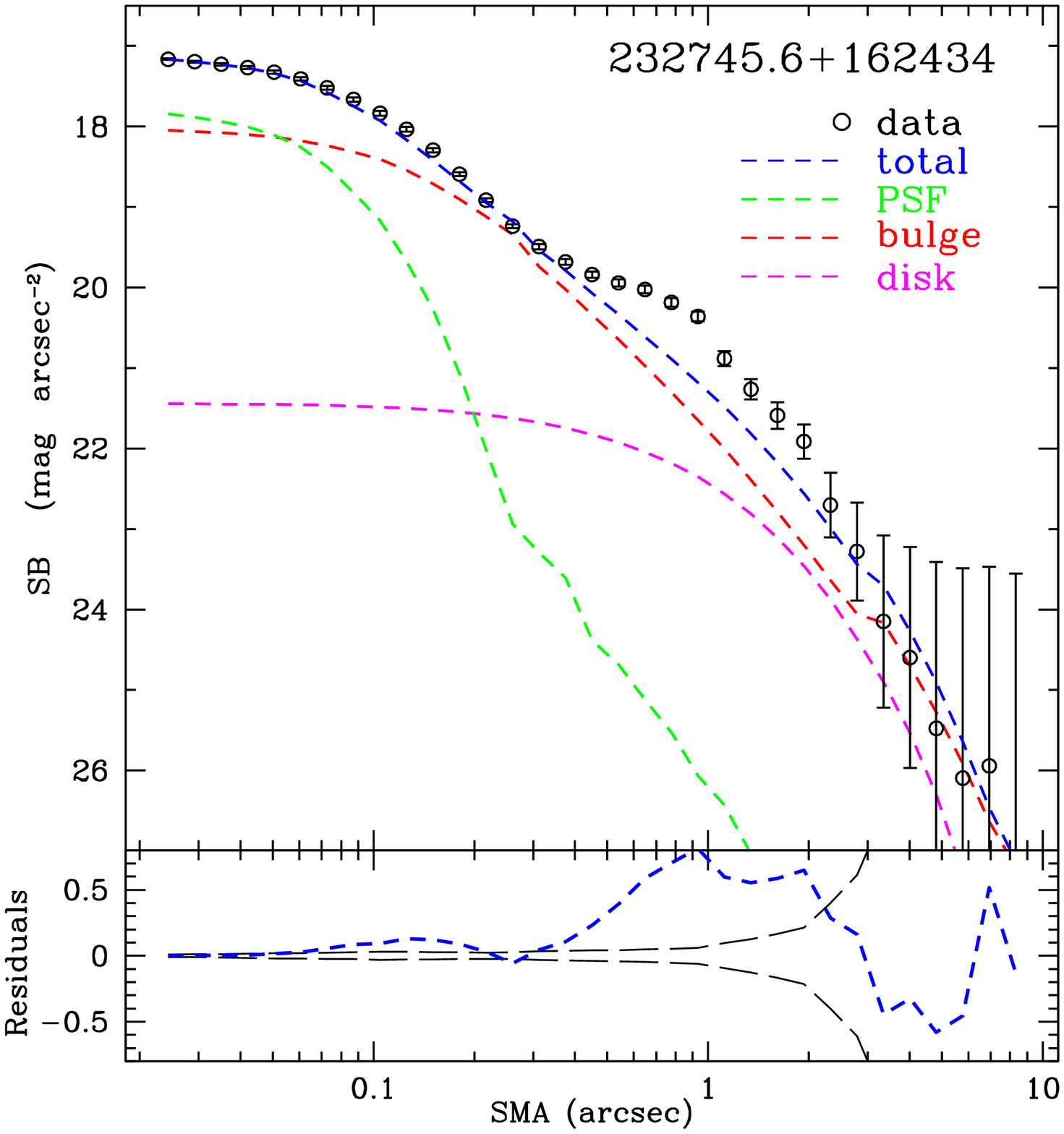}\\
\caption{One-dimensional surface brightness profiles of the $HST$ F814W images of the 2MASS QSOs, showing each of the components used to fit the objects with GALFIT. The black dashed lines in the residuals panels trace the one-sigma error bars in the observed data.}
\label{figure:radial}
\end{center}
\end{figure}
\twocolumn
\onecolumn
\begin{figure}
\begin{center}
\figurenum{6}
\epsscale{0.9}
\plotone{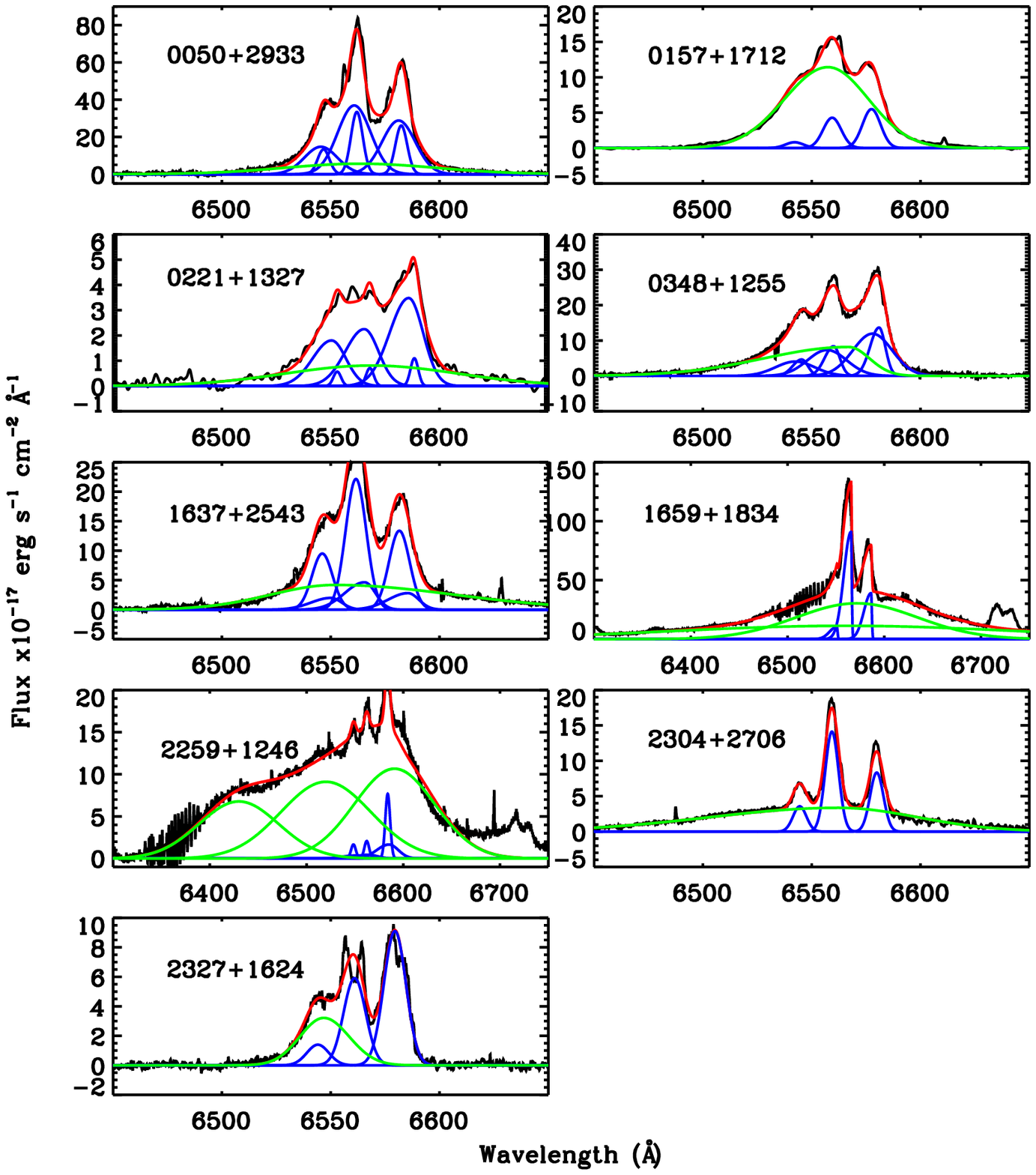}
\caption{Continuum-subtracted, rest-frame Keck ESI spectra (black trace) of the 2MASS QSOs in the \ha\ region.  The narrow line component models are plotted in blue, the broad emission lines in green, and the overall fit to the data in red.
}
\label{figure:lineprofiles}
\end{center}
\end{figure}
\twocolumn
\noindent
\citep[see][for details] {greene2005}, whereas \citet{mcgill2008} derive their relation for \ha\ line dispersions from only 19 objects.  Thus, we use the relation for FWHM$_{H\alpha}$ rather than that for line dispersions.  This choice does not affect our main results: if we were to use line dispersions rather than FWHM$_{H\alpha}$, the resulting \mbh\ would be, on average, 0.08 dex greater.

\begin{deluxetable}{clrcccc}
\tabletypesize{\small}
\tablecolumns{5}
\tablewidth{0pc}
\tablecaption{Black Hole Masses}
\tablehead{
\colhead {Object} & {QSO}    & Weighted & \colhead{FWHM$_{H\alpha}$}   & \colhead{E($B-V$)} & \colhead{$\lambda L_{5100}$} & \colhead{\mbh} \\
\colhead {ID} & {(2MASSi J)} & avg. \sig & \colhead{(km s$^{-1}$)}& \colhead{}          & \colhead{(10$^{44}$ erg s$^{-1}$)}& \colhead{(10$^{8}$ \msol)} }
\startdata
1&005055.7$+$293328 & $176_{-17}^{+14}$ &$2679^{+227}_{-175}$ & 1.64 & \phn2.52$\pm$1.24 & \phn$1.19^{+0.40}_{-0.55}$  \\
2&015721.0$+$171248 & $250_{-54}^{+54}$  &$2196^{+27}_{-32}$   & 2.13 & 18.74$\pm$2.05 & \phn$2.24^{+0.35}_{-0.72}$  \\
3&022150.6$+$132741 & $129_{-22}^{+22}$  &$4279^{+421}_{-492}$ & 1.76 & \phn1.78$\pm$0.24 & \phn$2.61^{+1.06}_{-1.01}$ \\
4&034857.6$+$125547 & $152_{-17}^{+17}$&$4213^{+355}_{-342}$ & 1.68 & \phn4.16$\pm$0.95 & \phn$3.93^{+1.47}_{-1.61}$ \\
5&163736.5$+$254302 & $129_{-12}^{+14}$  &$3178^{+371}_{-248}$ & 1.93 & 14.13$\pm$2.21 & $\phn4.15^{+1.87}_{-1.75}$  \\
6&165939.7$+$183436 & $159_{-25}^{+28}$ &$6742^{+3640}_{-142}$ & 0.39 & \phn0.94$\pm$0.24 & \phn$4.79^{+9.31}_{-1.38}$  \\
7&225902.5$+$124646 & $194_{-20}^{+21}$ &$4457^{+5543}_{-100}$ & 0.84 & \phn4.03$\pm$0.55 & \phn$4.34^{+27.9}_{-3.05}$  \\
8&230442.4$+$270616 & $127_{-7}^{+7\phn}$  &$6349^{+251}_{-265}$ & 0.99 & \phn1.71$\pm$0.47 & \phn$5.77^{+1.62}_{-2.02}$ \\
9&232745.6$+$162434 & $205_{-19}^{+19}$&$2341^{+416}_{-161}$ & 1.93 & 19.16$\pm$5.57    & \phn$2.59^{+1.55}_{-1.18}$ \\
\enddata
\label{mbh}
\end{deluxetable}

\subsubsection{Measuring $L_{5100}$}\label{L5100}

The AGN continuum flux at rest frame 5100 \AA\ cannot be measured directly from our ESI spectra for two primary reasons: (1) The continua of red QSOs suffer 
from significant extinction at short wavelengths, and this extinction needs 
to be properly measured.  (2) The spectrum in this wavelength region has
a large contribution from the host galaxy that also needs to be accounted for.

In order to obtain the unobscured QSO flux at 5100 \AA, we first 
measured the extinction for each object by modeling its spectrum.
As described in \S\ref{sigma}, we modeled the QSO continuum by 
including a polynomial (in $x = ln(\lambda)$ space) in the fitting procedure to obtain \sig.  The resulting
polynomial, $C(x) \times P(x)$, is then a good representation of the reddened QSO.  
In order to estimate E($B-V$), we compared the polynomial to the Sloan Digitized Sky Survey (SDSS) composite QSO spectrum \citep{vandenberk2001} reddened with a Small Magellanic Cloud (SMC) reddening law \citep{prevot1984,bouchet1985}, with E($B-V$) as a free parameter.
The values of E($B-V$) that we obtained using this procedure are listed 
in Table~\ref{mbh}.  Two thirds of our objects span a range of E($B-V$) values comparable to those found by \cite{urrutia2009} in a sample selected from the FIRST-2MASS Red Quasar Surveys \citep{glikman2007}, i.e., E($B-V$) $\lesssim 1.5$.  The remaining three objects, however have E($B-V$) $\sim 2$.  Typical errors in E($B-V$) derived from the fits are $\sim$0.1.

Whenever possible, we also estimated E($B-V$) using H$\alpha$/H$\beta$ ratios.  Naturally, the values for these ratios become more uncertain with 
increasing extinction, so we only estimated E($B-V$) for the seven objects for 
which we could obtain reliable ratios.   The resulting E($B-V$) values were
generally consistent with the values obtained from the fitting procedure,
although the two could differ by as much as E($B-V$)$\sim$0.3. We found no trend in the way the two values differed; the average E($B-V$) of the seven objects measured from the Balmer decrement was 1.49, compared to 1.45 as inferred from modeling the AGN continuum.  Since there is no a priori reason why the continuum and broad 
emission lines should suffer the same amount of extinction, we only used the Balmer decrement estimates as a rough comparison to make sure the reddening estimates derived from the model were reasonable.

As a further test, we subtracted the corresponding reddened SDSS composite 
spectrum from the spectrum of each of our targets to recover the host galaxy
spectrum.  There were some QSO features that were over- or under-subtracted
in the resulting host galaxy spectra.  This is caused by the mismatches 
between the SDSS composite spectrum and the intrinsic spectra of our QSOs. 
Indeed, the greatest source of uncertainty in the final values of $L_{5100}$
comes from such mismatches.
However, the overall spectral energy distribution (SED) of each host galaxy 
matched the shape of the SED that would be expected for a galaxy with the given set of stellar absorption features observed in that galaxy.  

Once we had determined the intrinsic reddening for each object, we scaled each continuum model, $C(x) \times P(x)$, to match the observed flux of
the corresponding QSO nucleus, as measured from $HST$ WFPC2 F814W images
using GALFIT (see Table~\ref{table:galfit}) and corrected for Galactic extinction using the values given by \citet{schlegel1998}.  
Thus we constructed continua of the reddened QSOs without galaxy contamination and taking into account flux
loses due to the slit.  Finally, we corrected these continua for intrinsic extinction (using the values listed in Table~\ref{mbh}) and measured the the flux of each QSO at rest-frame 5100 \AA. As mentioned above, the greatest source of uncertainty comes from measuring E($B-V$) by making the assumption that the intrinsic continuum shape of our sources is equal to that of the SDSS composite QSO.  The uncertainties for $\lambda L_{5100}$ listed in Table~\ref{mbh} include the uncertainty in the intrinsic power-law index, $\alpha_{\lambda}$ (conservatively assumed to be $\sim$0.05; \cite{vandenberk2001}) as well as the uncertainty in measuring E($B-V$) by the process we outline above.

The black hole masses derived from our measurements of FWHM$_{H\alpha}$ and $\lambda L_{5100}$ are listed in Table~\ref{mbh}.   For completion, we also repeat the values of \sig\ from Table~\ref{table:results} in Table~\ref{mbh}.  As mentioned before, using line dispersions rather than FWHM to estimate black hole masses results in masses that are, on average, 0.08 dex greater than those presented in this table.

\section{Results}\label{results}

\begin{figure}[tbh]
\figurenum{7}
\begin{center}
\epsscale{0.8}
\plotone{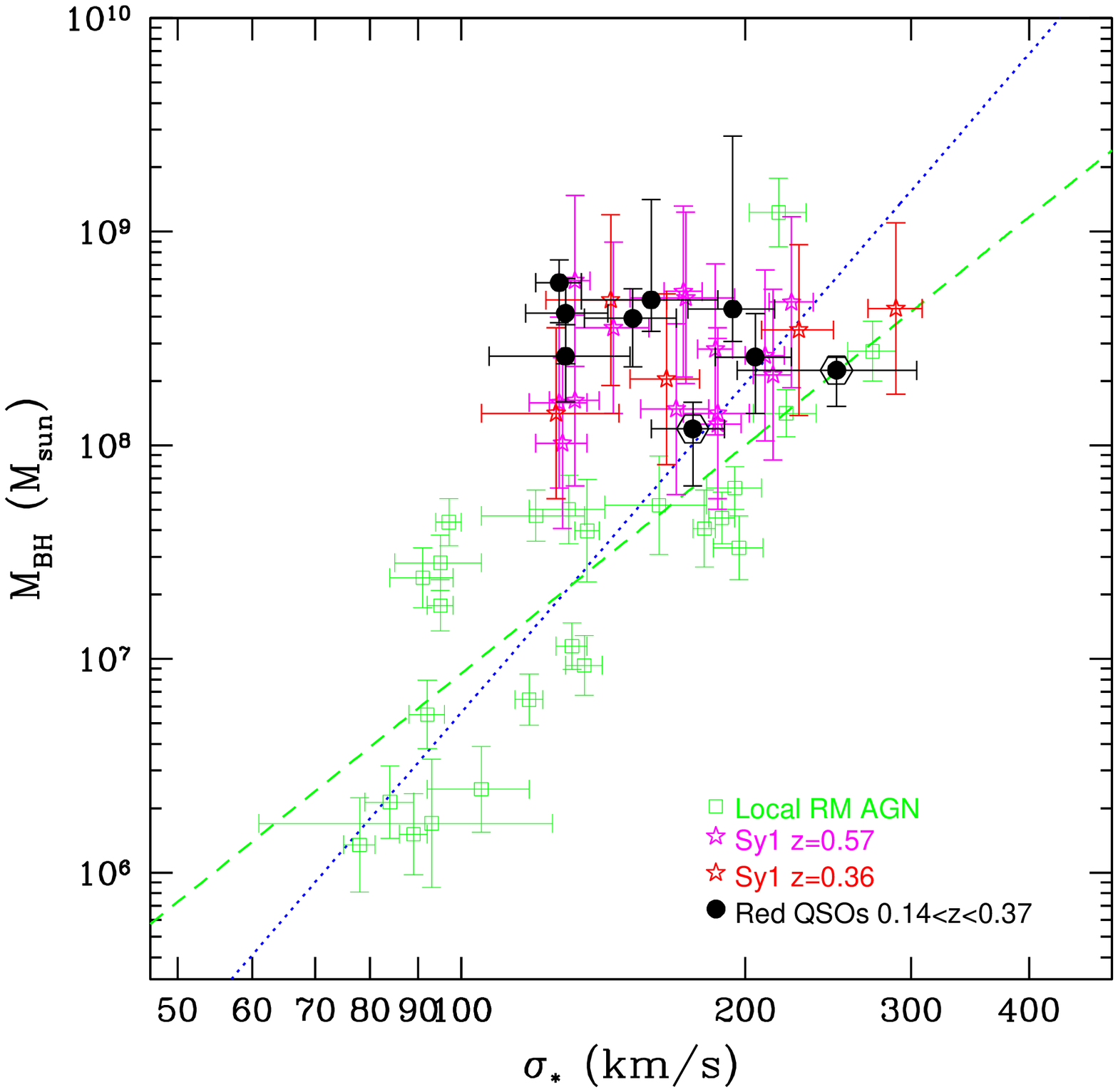}
\caption{Red 2MASS QSOs (black circles) in the \msig\ diagram.  
The green dash line is the \msig\ for local AGN relation measured by \cite{woo2010}.  We also plot the \msig\ relation for quiescent local galaxies measured by  \cite{mcconnell2011} as a blue dotted line.
Green squares are local AGN with reverberation mapped \mbh\ reported by \cite{woo2010}.  Stars denote the Seyfert galaxies at $z=0.36$ and $z=0.57$ from the work by Woo et al.~\citeyear{woo2006,woo2008} and \cite{treu2007}.  The two objects marked with hexagons do not follow the Faber-Jackson relation (see text for details)}
\label{msigmaplot}
\end{center}
\end{figure}

Our results are summarized in Fig.~\ref{msigmaplot}, where the 2MASS
red QSOs are plotted as black circles.  We also plot objects from \cite{woo2010}, which include the 24 local AGN that have published \sig\ and \mbh\ from reverberation mapping (green squares), as well as the \msig\ relation derived by \citeauthor{woo2010} from that sample (green dashed line).  For reference, the \msig\ relation for inactive local galaxies 
measured by \cite{mcconnell2011} is plotted as a blue dotted line.

The majority of the red QSOs are clearly above the local relations.  We have specifically compared our sample to AGN only to avoid any potential biases introduced by comparing active to quiescent galaxies \citep{lauer2007}, and we have used a virial coefficient consistent with that found by \citet{woo2010} from the reverberation mapped AGN sample.
\cite{woo2010} estimate an intrinsic scatter of 0.43 dex for their relation including all morphological types.  In contrast, our objects are, on average, 0.80 dex above the \citeauthor{woo2010} relation, with six out of nine of them having log(\mbh) above the intrinsic scatter.  
Comparison with the \citeauthor{mcconnell2011} relation for quiescent galaxies yields a similar, although smaller, discrepancy: our sample is, on average, 0.65 dex above the relation, and with five objects being above the intrinsic scatter.

When comparing our objects to those of \citet{woo2010}, it is pertinent to consider possible differences in the width of \ha\ when measured from single epoch (SE) spectra versus rms spectra from reverberation mapped objects.
\citet{park2012} find that H$\beta$ is broader in rms spectra than in SE spectra, although the difference between the two decreases as the intrinsic width of the broad lines increases.  \citeauthor{park2012} present a prescription to correct for this effect in relatively narrow-lined ($<$3000 \kms) objects.  To get an idea of how much this effect might impact our results, we can assume that the correction for H$\beta$ holds for \ha, and that it is a reasonable approximation for the objects in our sample with broader \ha\ as well.  With those assumptions, if we apply equation (8) of \citeauthor{park2012} to correct the FWHM$_{H\alpha}$ in Table~\ref{mbh}, we find that the black hole masses of the objects in our sample decrease by 0.13 dex on average.   This correction, however, is not enough to bring the objects to agreement with the local \msig\ relation, as they would still be, on average, 0.67 dex above the \citet{woo2010} relation.   Even if the correction for \ha\ and/or for objects with broader lines were greater, it is unlikely to be large enough to account for the offset we observe.

\begin{figure}[tbh]
\figurenum{8}
\begin{center}
\epsscale{1.1}
\plottwo{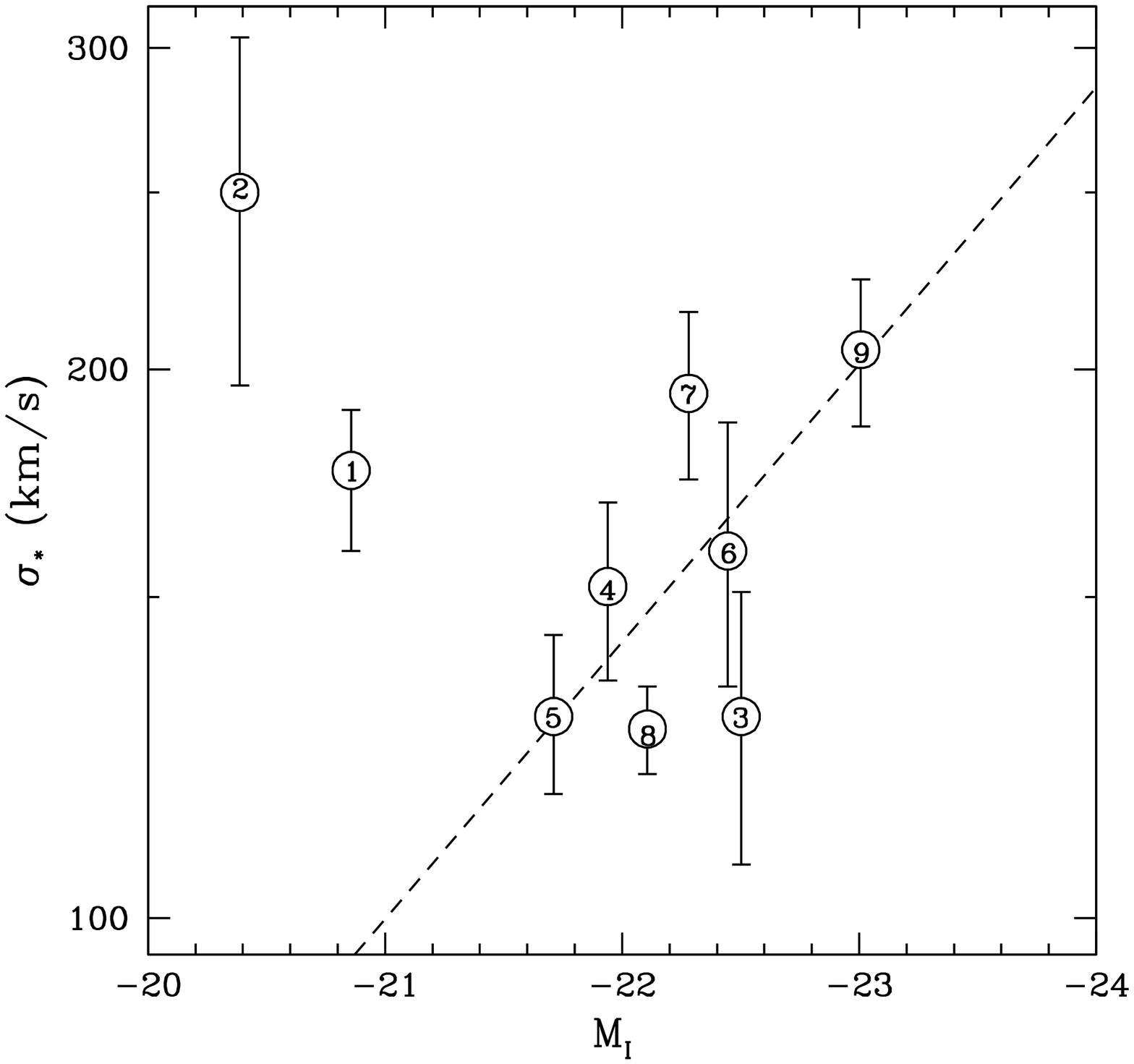}{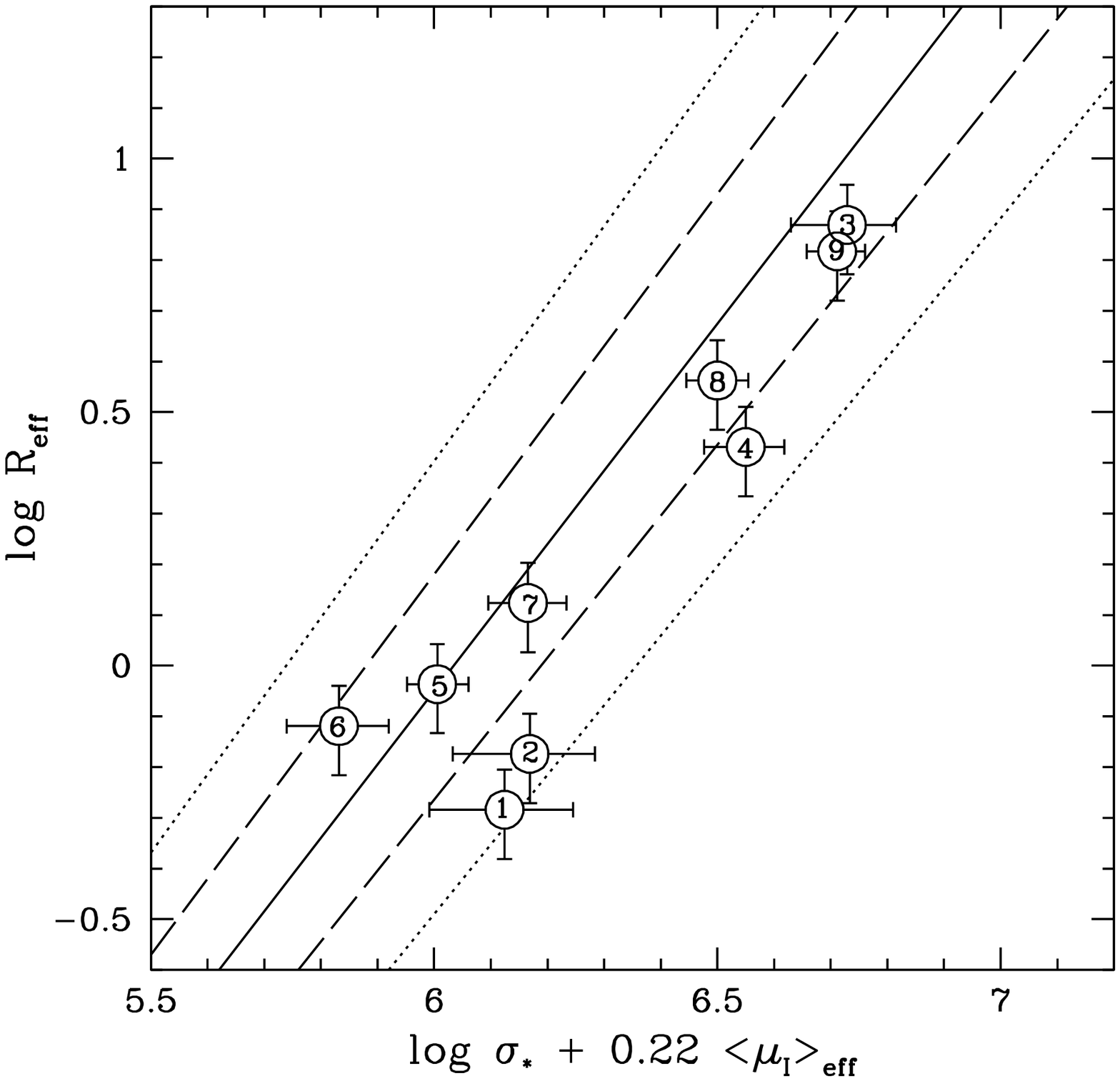}
\caption{{\bf Left:} Faber-Jackson relation for the bulge component of the 2MASS red QSO hosts.  The errors in M$_{\rm I}$ are smaller than the symbols (0.01$-$0.04 mag).  The numbers indicate the object ID number, as listed in the various tables.
The dashed line is the Faber-Jackson relation measured by \citet{nigoche2010} from a sample of $\sim$90,000 SDSS early-type galaxies in the luminosity range spanned by our sample in the SDSS $r-$band. Note that we have not accounted for potential wavelength-dependent differences in the slope. {\bf Right:} The bulges of the 2MASS red QSOs plotted on the Fundamental Plane. The solid line is the orthogonal fit to the Fundamental Plane from \citet{hyde2009} derived from $\sim$50,000 SDSS-DR6 early-type galaxies. The dashed and dotted lines correspond to the values that enclose 68\% and 95\% of the SDSS galaxies, respectively.}
\label{figure:fj}
\end{center}
\end{figure}

It is important to consider other potential biases that may be present in our sample.  First, the morphology of the host galaxies could introduce a bias. It has been shown that the precise form of the \msig\ relation has a dependence on the morphology of the galaxies in question \cite[e.g.,][]{graham2011}, and that pseudobulges do not follow the same relation \citep{kormendy2011}. 
As can be seen in Table~\ref{table:galfit}, every host galaxy in our sample has a bulge component that is best fit by a S{\'e}rsic of index $n\simeq4$.  Although most of the galaxies also appear to have a disk component, the majority (7/9) are clearly dominated by the bulge.  In order to investigate how the presence of the disk may affect our measurement of \sig\ in the bulges, we tested whether they follow the Faber-Jackson relation \citep{faber1976}. In the left panel of Fig.~\ref{figure:fj}, we plot velocity dispersions vs.\ absolute Cousins I magnitudes of the bulge components of the galaxies.  The latter were obtained using the output magnitudes from GALFIT given in Table~\ref{table:galfit} corrected for Galactic reddening, applying k-corrections for the spectral type of each galaxy and passive evolution using the on-line calculator from \citet{vandokkum2001}.  
The dashed line in the left panel of Fig.~\ref{figure:fj} is the Faber-Jackson relation from \citet{nigoche2010}.
\citeauthor{nigoche2010} use a sample of $\sim$90,000 SDSS-DR7 early-type galaxies to calculate the Faber-Jackson parameters (i.e., A and B in log~\sig = A - BM$_{r}$) for different magnitude ranges.  Our objects span a magnitude range of $-20.6<M_{I}<-23.4$, or $-19.8<M_{r}<-22.6$ \cite[AB magnitudes, using the transformation equations given by][]{jester2005}.  In this magnitude range, \citet{nigoche2010} find A = 0.152$\pm$0.009 and B = $-$0.976$\pm$0.193.  Note that the relation of \citeauthor{nigoche2010} is measured in the SDSS r-filter, and we have simply transformed it to Cousins-I, without accounting for potential wavelength-dependent differences in the slope. 

Two host galaxies fall clearly off the relation: those of 005055.7+293328 and 015721.0+171248 (objects \# 1 and 2, respectively), both of which have disks with luminosities comparable to, or greater than, their corresponding bulges.  As described in \S~\ref{morphologies}, the host galaxy of 005055.7+293328 has prominent nearly edge-on disk with spiral arms and a dust lane that goes through the bulge.  To mimic the dust lane, we fit this object with two separate components, so we almost certainly underestimate the flux contained in the bulge. \citet{marble2003} fit a single component to the host galaxy of this object and find a M$_{I}$ = $-$22.3.  This magnitude would bring the position of the object to agreement with the Faber-Jackson relation.  The other object, 015721.0+171248, is the only one in the sample that has a disk that is more luminous than the bulge.  The rotating component from a dynamically cold disk can bias \sig\ toward higher values (when the disk is edge-on) or lower values \citep[when the disk is face-on; see e.g.,][] {bennert2011a}.  Although the disk of 015721.0+171248 has a lower inclination than that of 005055.7+293328, its velocity dispersion may still be somewhat increased by the dynamically cold component.   
More importantly, 015721.0+171248 is the only object for which there is significant QSO contamination in the stellar absorption line spectrum, so that we were not able to obtain a reliable fit of \sig\ for the red region. 
We have marked both of these objects with hexagons in Figs.~\ref{msigmaplot} and \ref{deltambh} to indicate that the \sig\ measured in these objects may not be representative of normal bulges.  Interestingly, these are the two objects that fall closest to the local AGN \msig\ relation, and removing them from the analysis only accentuates the offset of the red QSO sample with respect to the local relation.

We also consider the position that the bulges in our sample occupy on the Fundamental Plane \cite[FP;][]{djorgovski1987}.  We obtain $<\mu>_{eff}$ as an output parameter from GALFIT, applying k-corrections and passive evolution to $z=0$ as above, as well as correcting from cosmological dimming.  We compare it to the FP relation derived by \citet{hyde2009} from a sample of $\sim$50,000 SDSS-DR4 (with parameters from DR6) early-type galaxies at $0<z<0.35$.  Here we use their orthogonal fit to the plane in the SDSS $i-$band and we have transformed it to Cousins-I, using the corrections prescribed by \citet{rothberg2012}.   The solid line in Fig.~\ref{figure:fj} (right panel) corresponds to the coefficients given in Table 2 of \citet{hyde2009}, and the dashed and dotted lines correspond to the values that enclose 68\% and 95\% of their SDSS early-type galaxies, respectively.  While, on average, our objects fall below the FP, they are consistent with the scatter in the relation.  The two objects that are off the Faber-Jackson relation also appear to be the ones with the largest offset from the FP.

Besides the effects of a potential dynamically cold component on \sig, we need to consider the possibility that we are measuring \sig\ in systems that are not dynamically relaxed.  At least a third of the host galaxies in our sample show clear signs of interactions (Figs.~\ref{figure:galfit}, \ref{figure:galfit_b}, \ref{figure:galfit_c}).
Many of the numerical simulations that are successful at reproducing the \msig\ relation presuppose that AGN activity is triggered by mergers \cite[e.g.,][]{robertson2006a,robertson2006b,hopkins2006}.  However, they predict the value \sig\ after the system has reached dynamical equilibrium.  To investigate the evolution of \sig\ during mergers, \cite{stickley2012} use N-body simulations of merging galaxies to perform a flux-weighted measurement of \sig\ through a diffraction slit.  They find that, typically, \sig\ tends to increase right after the first passage, followed by a period of oscillation about the value that \sig\ eventually reaches once the system comes to dynamical equilibrium.  Thus, the observed value of \sig\ in our objects could potentially be different from their corresponding equilibrium value, depending on the precise merger stage at which we happen to be observing them.   However, \citeauthor{stickley2012} estimate that, even in extreme cases, the observed value of \sig\ should fall between 70\% and 200\% of the quiescent value, and the probability that the observed value of \sig\ will actually be far from the equilibrium value is rather small.  Therefore, even in the highly unlikely case that our entire sample is caught during an oscillation stage when \sig\ has a lower value, the predicted quiescent value would not be high enough to bring our sample to agreement with the local \msig\ relation.    Moreover, using more sophisticated high-temporal resolution numerical simulations, Stickley and Canalizo (in preparation) find that the time at which the value of \sig\ shows significant deviations from the equilibrium value preceeds the period of measurable accretion onto the black hole (i.e., the time when the object appears as a QSO).  

Another possibility is that dust in the host galaxy may be obscuring a fraction of the stars and this may result in a biased measurement of \sig.   With the exception of 005055.7+293328, we find no clear evidence for significant amounts of dust in the host galaxies of the red QSOs.  Their spectra show that, while the nucleus suffers from strong extinction, the host galaxies are not significantly reddened.  As mentioned in \S~\ref{L5100}, after subtracting a reddened QSO spectrum, the overall SED of each host galaxy matches the shape of the SED that would be expected from the observed stellar absorption features.  This indicates that there is no severe reddening in the host galaxies.  Moreover, if the bulges of these galaxies were significantly reddened, they would not fall on the Faber-Jackson or FP relations (see Fig.~\ref{figure:fj}). However, we cannot rule out the possibility that a modest amount of dust reddenning (E($B-V$)$\lesssim$0.1) could be present in the hosts.   Therefore, it is important to consider the effects that dust in the host galaxy, particularly if located close to the central regions of the bulge, would have on the measurement of \sig.   The simulations of \cite{stickley2012} include a toy model for dust consisting of a slab of gray attenuating material, and show that a lower value of \sig\ can be measured if the light from stars closer to the central regions is more extinguished than that of stars at larger radii.  The specific toy model used by \citeauthor{stickley2012} leads to a decrease on the measured value of \sig\ $\lesssim$15\%.  Correcting for this factor would not bring the red 2MASS QSOs to the local \msig\ relation. While more realistic models for dust attenuation are needed to truly quantify the effect of dust on the measurement of \sig, we repeat that there is no evidence indicating that our host galaxies suffer from significant dust extinction.

\begin{figure}
\figurenum{9}
\begin{center}
\epsscale{0.8}
\plotone{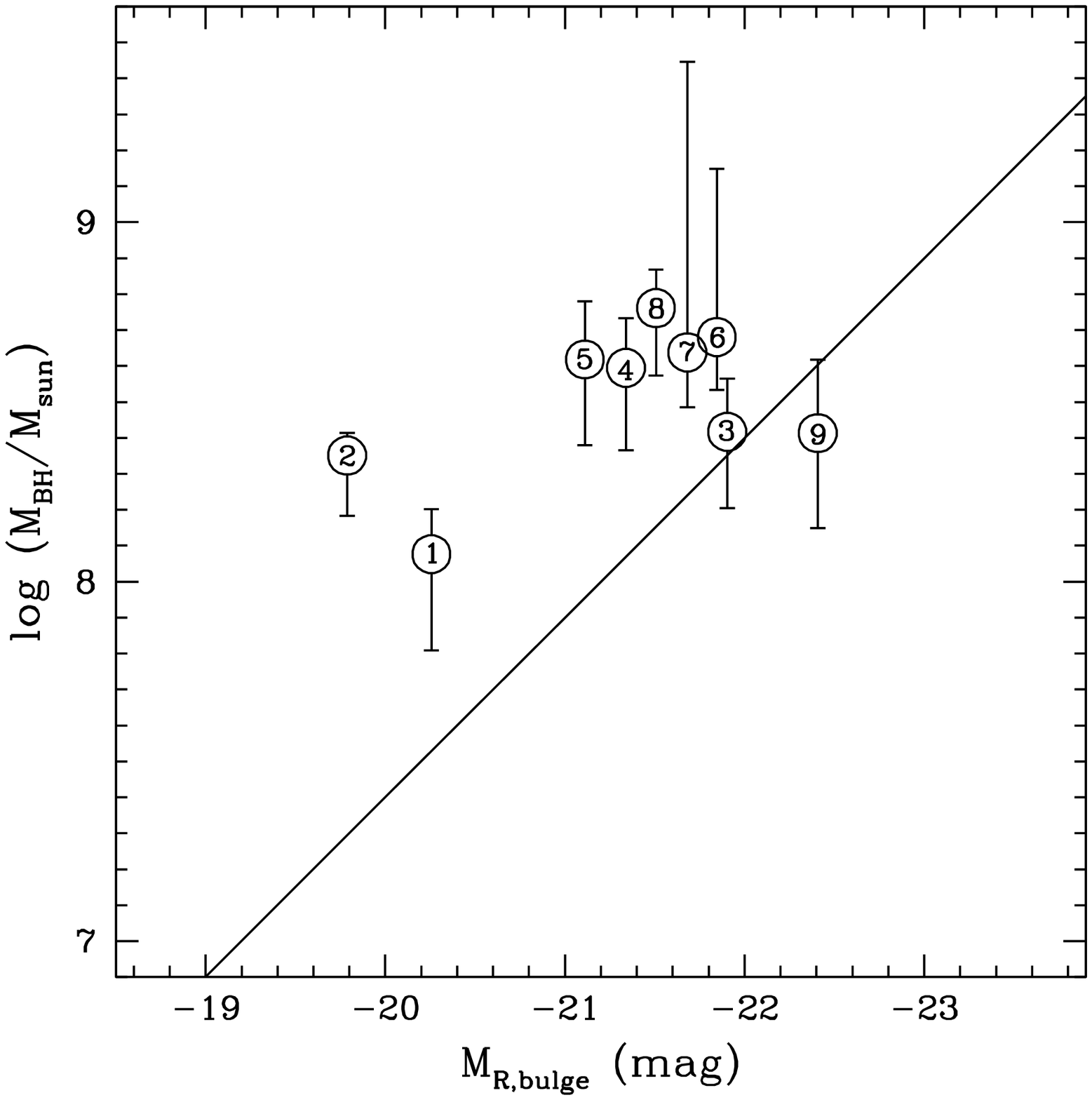}
\caption{Red 2MASS QSOs in the \mbh - L$_{\rm bulge}$ diagram. The solid line is the local relation for inactive galaxies from \citet{bettoni2003}, as reported by \citet{kim2008}. The numbers indicate the object ID number, as listed in the various tables.  The errors in M$_{\rm R}$ are, on average, 0.1 mag.  The magnitude of object 1 is likely to be underestimated due to the presence of a dust lane (see text for details). Note that objects 1 and 2 do not fall on the Faber-Jackson relation.}
\label{figure:mlbul}
\end{center}
\end{figure}

Finally, we consider the bias that may be introduced from the sample selection.  The objects were selected based on their near-infrared colors.  The selection criterion that objects must have $J-K_{s}>2.0$ implies that their SEDs will be dominated by the AGN in the near-infrared; if they were dominated by the host galaxy, their colors would be bluer than this.   Thus, the host galaxies of the most reddened AGN could possibly be biased toward lower luminosity (and lower mass).  We investigate this potential bias by looking at the offset of the objects as a function of near-infrared colors and reddening.  First, we plot our objects on the black hole mass - bulge luminosity relation, \mbh - L$_{\rm bulge}$, in Fig.~\ref{figure:mlbul}, where we have transformed the bulge magnitudes, M$_{I}$ from Fig.~\ref{figure:fj} to M$_{R}$ assuming an average $R-I$ = 0.6 for the sample.   The solid line in this figure is the local relation for inactive galaxies from \citet{bettoni2003}, as reported by \citet{kim2008}.  Not surprisingly, our objects have overmassive BHs with respect to the local relation: they are, on average, 0.89 dex above the relation.    Note that two of the objects that have the largest offsets with respect to the relation (objects 1 and 2) are also the two objects that do not follow the Faber-Jackson relation.  If we do not include these two objects, then the sample is, on average, 0.69 dex above the relation.

\begin{figure}
\figurenum{10}
\begin{center}
\epsscale{1.1}
\plottwo{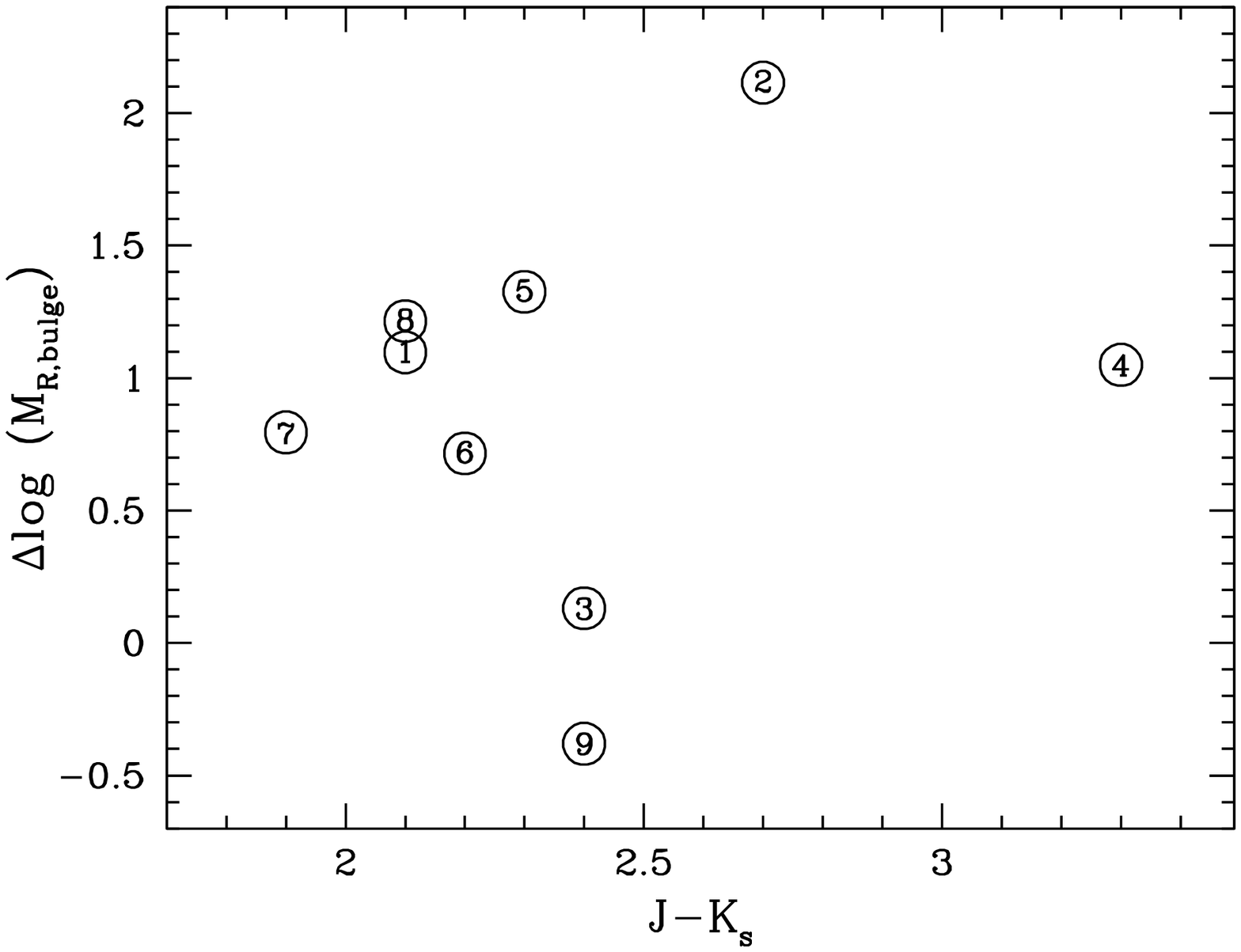}{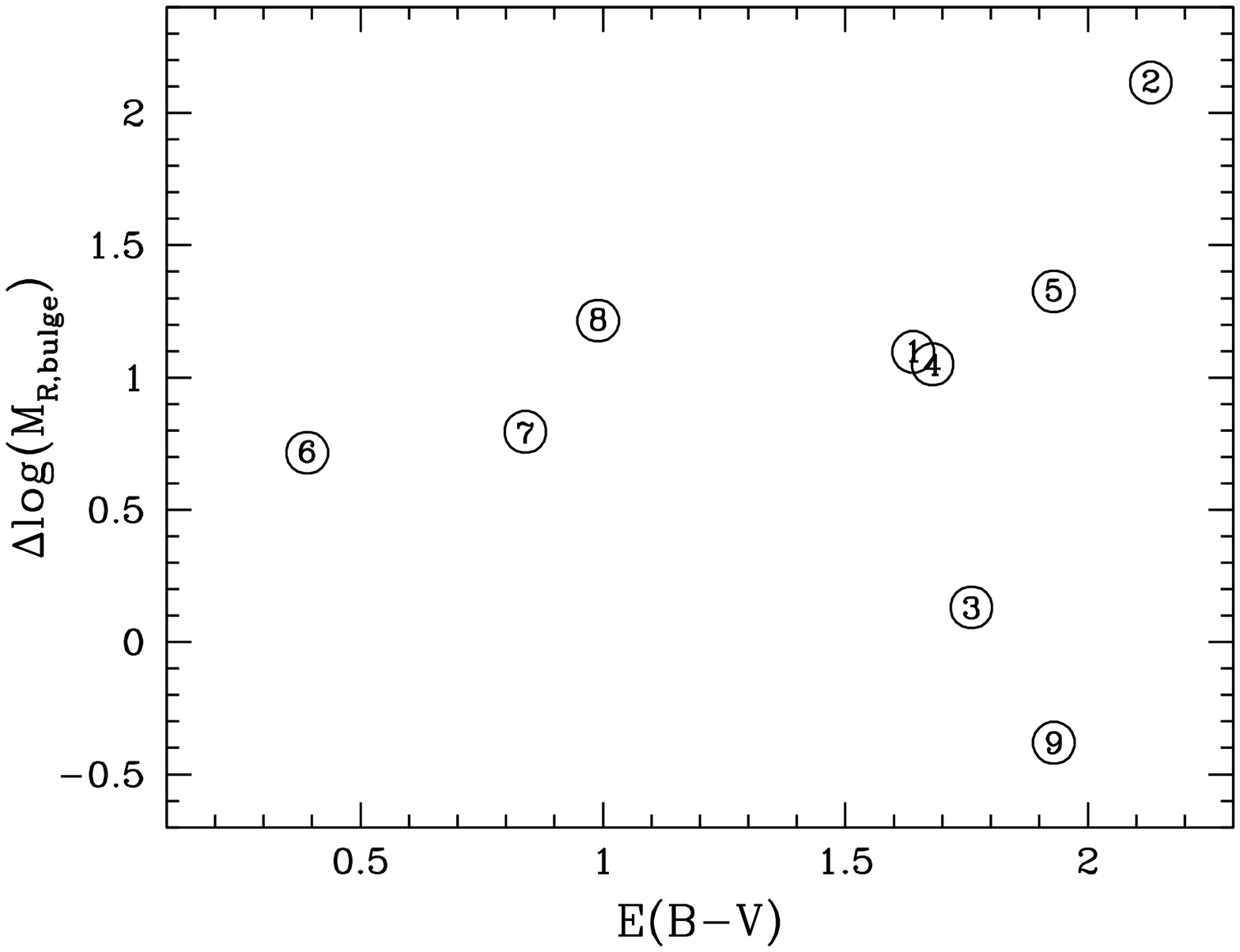}
\caption{Offset from the \mbh - $L_{\rm bul}$ relation as a function of $J-K_{s}$ color (left) and reddening (right).
The numbers indicate the object ID number, as listed in the various tables.  Note that objects 1 and 2 do not fall on the Faber-Jackson relation.}
\label{figure:colors}
\end{center}
\end{figure}

If there were indeed a bias favoring less luminous galaxies for more reddened objects, we should see a trend of an increasing offset from the \mbh - $L_{bulge}$ relation, $\Delta$M$_{R,bulge}$, with increasing $J-K_{s}$ colors and with increasing reddening, E($B-V$).   Figure~\ref{figure:colors} shows no such trends and, in fact, some of the most reddened objects also have the smallest offsets in magnitude.   Therefore, we do not see evidence for a bias introduced by the color selection of our sample.

\section{Discussion}\label{discussion}

Our sample appears to have a true offset with respect to the local \msig\ relation, in the sense that the black holes in red QSOs are overmassive compared to their host galaxies.   
\cite{woo2006,woo2008} and \cite{treu2007} find similar offsets for samples of Seyfert 1 galaxies at $z\sim0.36$ and $z\sim0.57$; their results are plotted along with ours in Fig.~\ref{msigmaplot}.  
\cite{woo2008} investigate the \msig\ relation at three different redshift bins ($<$0.1, 0.36, and 0.57), and conclude that there is evolution at the 95\% confidence level.  This is at least in qualitative agreement with several studies of the \mbh - \mbulge\ relation at higher redshifts. Using 51 quasar hosts (mostly lensed), \cite{peng2006a,peng2006b,peng2006c} find evidence that the ratio \mbh/\mbulge\ increases above $z=1$.  
\cite{bennert2011b} find similar results in a sample of $1<z<2$ X-ray selected, broad-line AGN by deriving stellar masses of the bulge component from multi-filter surface photometry in $HST$ images.  The implication from these and other studies \cite[e.g.,][]{decarli2010} is that at $z \sim 2$ host bulges were undermassive relative to their BHs (compared to today), suggesting that BHs grew first and bulges have been playing ``catch-up.''  

\begin{figure}[tbh]
\figurenum{11}
\begin{center}
\epsscale{1.1}
\plottwo{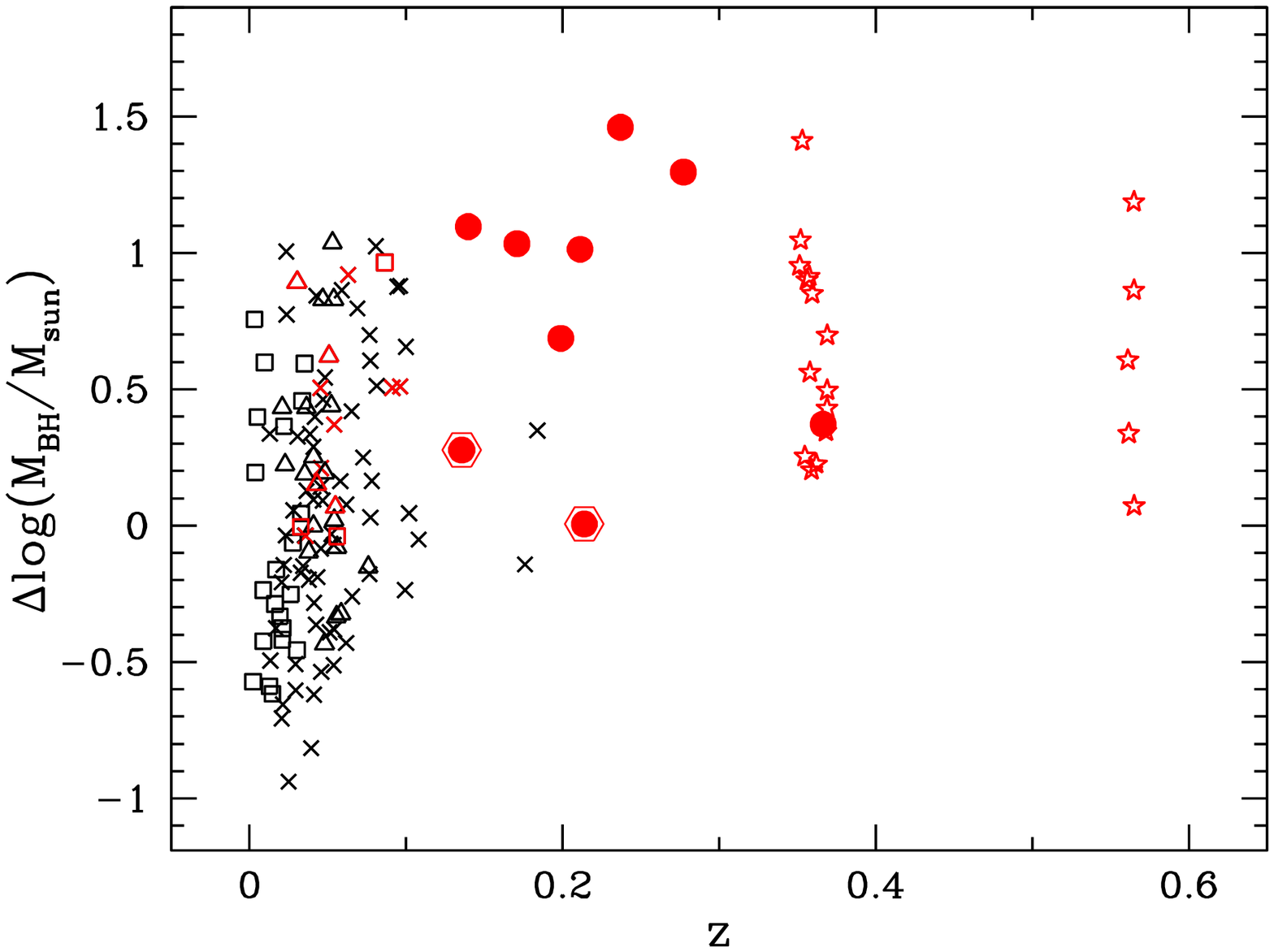}{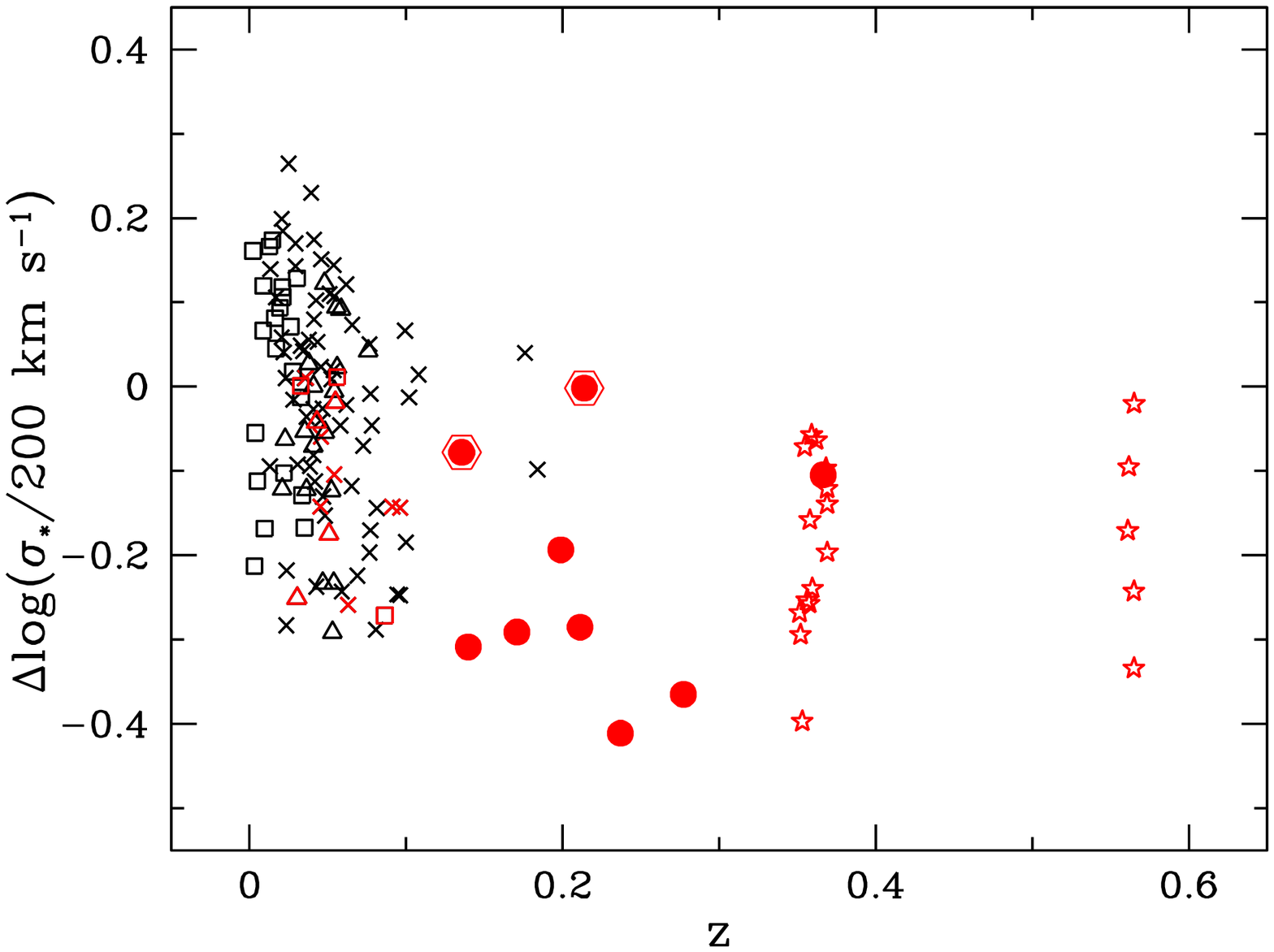}
\caption{{\bf Left Panel:} A representation of the offset in \mbh\ with respect to the local AGN \msig\ as a function of redshift, after \cite{woo2008}.  The offsets are with respect to the \cite{woo2010} \msig\ relation for local AGN.
Open stars are the samples of \cite{woo2006,woo2008}, crosses are the AGN from \cite{greene2006a}, open squares are reverberation mapped AGN \citep{woo2010}, and open triangles are AGN from \cite{bennert2011a}.  The 2MASS QSOs (this work) are plotted as filled circles, bridging the gap between local samples and higher redshift samples. The two objects marked with hexagons do not follow the Faber-Jackson relation in Fig.~\ref{figure:fj}.  Red symbols denote objects with log(\mbh/\msol) $> 8$. {\bf Right Panel:} Same as left panel, but plotting the offset in \sig.
}
\label{deltambh}
\end{center}
\end{figure}

While our results appear to be in line with all these studies, they also present additional challenges to the evolutionary picture.  \cite{woo2008} find a potential trend in the evolution of \msig, if indeed the higher redshift samples are the direct progenitors of the local AGN.  They show this clearly in their Fig.~3, where they plot the offset in \mbh\ with respect to the local, quiescent sample of \cite{tremaine2002}.  We repeat this plot, including our objects, in Fig.~\ref{deltambh}, where we have calculated the offsets with respect to the local AGN \msig\ relation from reverberation mapped objects \citep{woo2010}, and we also include the local AGN samples of \citet{bennert2011a} and \citet{greene2006a}. The BH masses in the \citet{greene2006a} sample have been increased by log(1.8) = 0.255 to account for the difference in the virial coefficient used by \citeauthor{greene2006a} compared to that used in the other studies.  Thus, all the BH masses in Fig.~\ref{deltambh} assume a virial coefficient of $f\sim5.5$ \citep{onken2004,woo2010}.

Taken at face value, Fig.~\ref{deltambh} would indicate that there is a steep increase in the offset between redshifts of 0.1 and 0.2, and that most of the evolution has occurred in the last 3 Gyrs.  Alternatively, in the context of models that find a non-causal origin for the scaling relations \citep[e.g.,][]{jahnke2011}, most of the mass existing in the disks of these objects has been transferred to the bulge through mergers in the last 3 Gyrs.  This would also be consistent with our result showing that the two objects with the most significant disk components are also the two objects with the largest offsets in the \mbh - L$_{\rm bulge}$ relation (Fig.~\ref{figure:mlbul}).
However, the comparison between local and higher redshift AGN samples is not straightforward, as the latter have, on average, significantly higher \mbh\ than their local counterparts.   Every object at $z>0.2$ has \mbh\ $> 10^{8}$ \msol, whereas there is a dearth of objects in this mass range at lower redshifts.   In Fig.~\ref{deltambh}, we plot all the objects with \mbh\ $> 10^{8}$ \msol\ with red symbols.  If we consider the red objects only, the trend with redshift becomes less significant.  The red objects at $z<0.1$ have an average offset of 0.40 dex, compared to 0.80 dex in our sample, and 0.66 dex and 0.61 dex, respectively, in the Seyfert 1 samples at $z=0.36$ and $z=0.57$ of \citet{woo2006,woo2008}.

The difference in \mbh\ range between the local and the higher-$z$ samples is due, at least in part, to the fact that they are drawn from flux-limited samples.  In these samples, less luminous objects are not detected at higher redshifts and, since AGN luminosity depends partially on \mbh, objects with lower \mbh\ are less likely to be picked.
\citet{lauer2007} describe a potential bias that may be present in samples of AGN at the highest \mbh.
They show that a bias may arise from the conspiratory nature of a steep luminosity function and intrinsic scatter in the \msig\ relation, leading to a selection that favors higher mass BHs over lower mass BHs. Thus studies targeting luminous or massive AGN may find an offset from the local relation that is only due to this selection bias.  On the other hand, \citet{treu2007} run Monte Carlo simulations to investigate the effects on a hypothetical selection function log(\mbh/\msol) $> 7.9$, which applies to the objects of \citet{woo2006,woo2008} as well as our own.  They compare the measured offset to a simulated input offset with respect to the local \msig\ and find that the bias is negligible, unless the intrinsic scatter at higher redshifts increases dramatically (of the order of 1 dex).   However, the fact that we find an offset from the local relation in all of the objects with \mbh\ $> 10^{8}$ \msol, both in the local and the non-local universe, may indicate that there is still some bias at play.

\begin{figure}
\figurenum{12}
\begin{center}
\epsscale{0.8}
\plotone{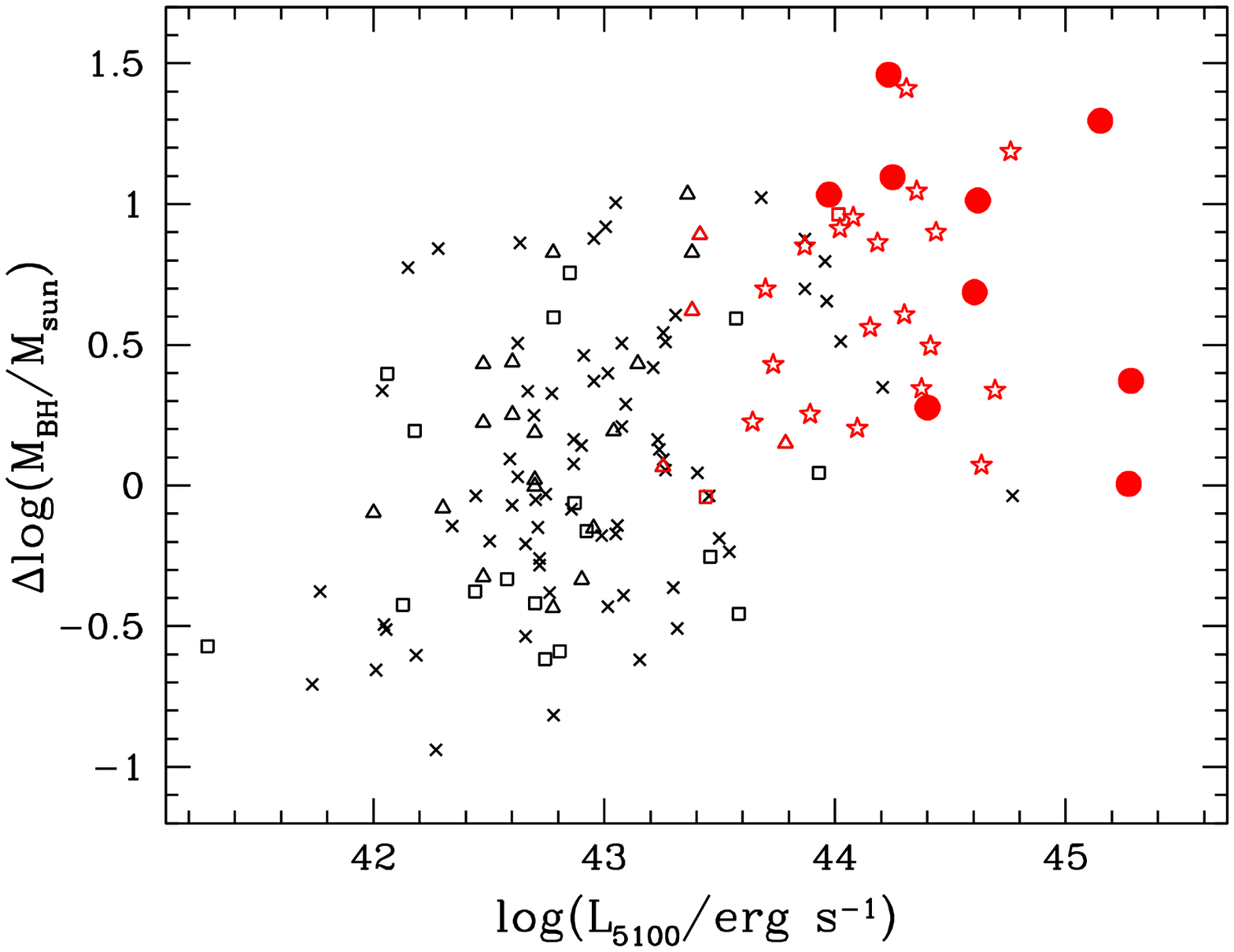}
\caption{Offset from the \msig\ relation as a function of AGN luminosity at 5100 \AA.  The 2MASS QSOs (this work) are plotted as filled circles. Open stars are the non-local samples of \citet{woo2006,woo2008}, crosses are local AGN from \cite{greene2006a}, open squares are local reverberation mapped AGN \citep{woo2010}, and open triangles are local AGN from \cite{bennert2011a}.  Red symbols denote objects with log(\mbh/\msol) $> 8$. 
}
\label{figure:luminosity}
\end{center}
\end{figure}

\begin{figure}
\figurenum{13}
\begin{center}
\epsscale{0.8}
\plotone{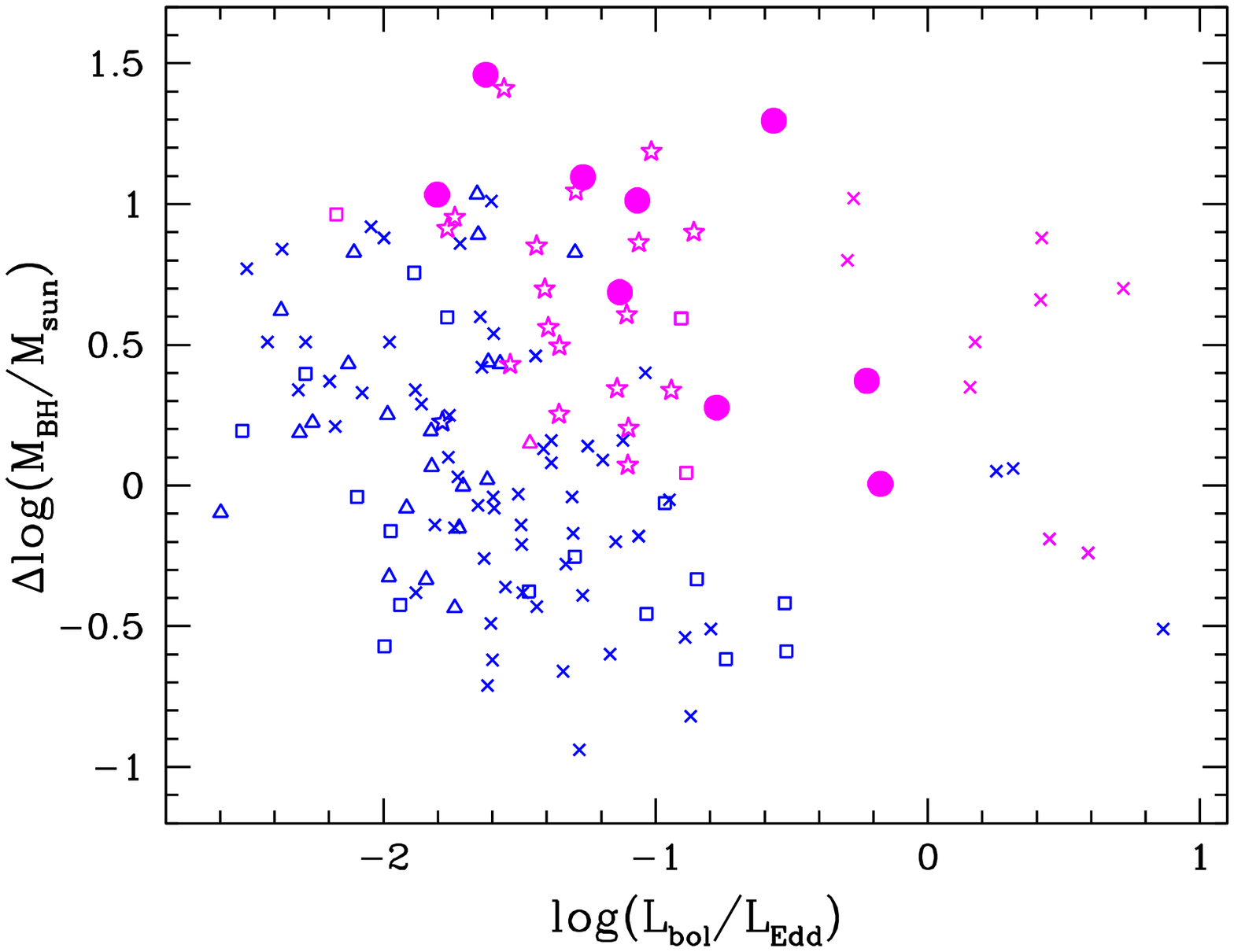}
\caption{Offset from the \msig\ relation as a function of Eddington ratio.  The 2MASS QSOs (this work) are plotted as filled circles. Open stars are the samples of \citet{woo2006,woo2008}, crosses are AGN from \cite{greene2006a}, open squares are reverberation mapped AGN \citep{woo2010}, and open triangles are AGN from \cite{bennert2011a}.  Blue symbols are objects with log (L$_{5100}$/erg s$^{-1}$) $<$ 43.5 and pink symbols are the objects with higher luminosities.}
\label{figure:eddington}
\end{center}
\end{figure}

In order to further assess potential biases, we examine the dependence of the \msig\ offset on AGN luminosities and Eddington ratios (Figs.~\ref{figure:luminosity} and \ref{figure:eddington}).   For the reverberation mapped AGN of \citet{woo2010}, we use the values of L$_{5100}$ published by \citet{park2012}, \citet{Bentz2006}, \citet{kaspi2000}, and \citet{denney2006}.   For consistency, we transform L$_{H\alpha}$ in the sample of \citet{greene2006a} to L$_{5100}$, using the relations given by \citet{greene2005}.   The resulting plot is shown in Fig.~\ref{figure:luminosity}.    We see a striking trend: The offset increases with luminosity, with virtually all the objects with log(L$_{5100}$/erg s$^{-1}$) $>43.6$ being above the relation, including those with \mbh\ $<$ 10$^{8}$ \msol.   Clearly, the majority (two thirds) of high luminosity AGN are in the non-local universe, and this could be driving the trend.  However, we note that all of the high-luminosity objects in the local universe follow this trend as well.

One possible way to explain this trend might be if the radius-luminosity ($R_{BLR}-L$) relation somehow changed dramatically at high luminosities, if, for example, a point is reached where the broad line region is matter-bound so that the radius-luminosity relation would flatten.  A different relation might also be expected if the SEDs of high luminosity objects are different from those of lower luminosity objects. The data points in Fig.~5 of \citet{bentz2009} suggest that, indeed, some flattening of the $R_{BLR}-L$ relation may occur at the highest luminosities.  However, even if that were the case, the potential flattening occurs at luminosities significantly higher than log(L$_{5100}$/erg s$^{-1}$) $= 43.6$.  In fact, if we were to fit only the objects in the luminosity range of $43.6 <$ log(L$_{5100}$/erg s$^{-1}$) $<$ 45.3 (corresponding to the luminous objects of Fig.~\ref{figure:luminosity}), we would obtain a steeper relation, closer to the relation of \citet{kaspi2000}, and resulting in somewhat higher black hole masses in this range.  Thus, variations in the $R_{BLR}-L$ relation cannot account for the trend we see in Fig.~\ref{figure:luminosity}.   

Using semi-analytical models of galaxy evolution, \citet{portinari2011} find that more luminous QSOs tend to trace over-massive BHs with respect to the intrinsic relation \cite[similar to the bias described by][]{lauer2007}.  However, for objects at $z<0.5$, their simulations predict that the parameters of QSO host galaxies closely follow the distribution of those of inactive galaxies, with very small offsets ($\sim$0.1 dex) in host galaxy mass.

To measure Eddington ratios we obtain rough bolometric luminosities by assuming L$_{\rm bol}$ = 10 L$_{5100}$ \citep{woo2002}.  The plot of $\Delta$\mbh\ vs.\ L$_{\rm bol}$/L$_{\rm Edd}$ (Fig.~\ref{figure:eddington}) shows large scatter and no correlations.   However, since Fig.~\ref{figure:luminosity} shows that there may be an offset in $\Delta$\mbh\ at high AGN luminosities, it may be instructive to consider separately objects of lower luminosity (log (L$_{5100}$/erg s$^{-1}$) $<$ 43.5; blue symbols in Fig.~\ref{figure:eddington}) and those of higher luminosity (pink symbols).  A trend appears to surface: objects with low L$_{\rm bol}$/L$_{\rm Edd}$ have more positive offsets with respect to \msig\ than those with higher L$_{\rm bol}$/L$_{\rm Edd}$.  We caution that the trends may be largely driven by the fact that we are dividing the objects in bins according AGN luminosity, e.g., a highly luminous AGN can have low \mbh\ only if it has a high accretion rate.  Such trends of increasing L$_{\rm bol}$/L$_{\rm Edd}$ with decreasing \mbh\ have been observed in large flux-limited samples of AGN \cite[e.g.,][]{netzer2007}.  However, we note that the trends we see in Fig.~\ref{figure:eddington} are in the offset with respect to \msig\ and not just in \mbh.

A similar trend with respect to \mbh-L$_{\rm bulge}$ was observed by \citet{kim2008} in a study of 45 $z\lesssim 0.35$ type 1 AGN.  They find that, at a given bulge luminosity, objects with high L$_{\rm bol}$/L$_{\rm Edd}$ have lower \mbh.  
They suggest that this may be explained in terms of the growth phase of BH: If L$_{\rm bol}$/L$_{\rm Edd}$ decreases as a function of time during a black hole growth episode, then BH with the highest L$_{\rm bol}$/L$_{\rm Edd}$ have more ``catching up to do'' than those with lower L$_{\rm bol}$/L$_{\rm Edd}$.  We searched for this trend by dividing the objects of Fig.~\ref{figure:eddington} in bins of stellar velocity dispersion (as a proxy for bulge luminosity; not shown in Fig.~\ref{figure:eddington}).   Given the large scatter, we found no significant trends or correlations.   We also searched for this trend in our sample by plotting $\Delta$M$_{R,bulge}$ (from Fig.~\ref{figure:mlbul}) against L$_{\rm bol}$/L$_{\rm Edd}$.  Once again, we found a large scatter and no significant trends.

The combination of the potential biases in these samples with the trends that we have uncovered in this study prevents us from making any definitive statements on whether we see evolution with redshift in the \msig\ relation.
Clearly, better statistics are needed to disentangle the dependence of \msig\ on mass, luminosity, accretion rates, and redshift.  Studying samples with a wider range of these properties could help us to better understand the scatter and the different biases.
For local AGN, \cite{bennert2011a} indicate a future study of $\sim$75 additional AGN with \mbh\ $> 10^{7}$ \msol\ that can be used to probe the local \msig\ relation at the highest \mbh\ and, likely, at high luminosities.   In the non-local universe, we are conducting a study of additional red QSOs as well as other QSOs that are similarly suited for the study of \msig, including objects in lower \mbh\ and lower luminosity ranges.  We will present results from objects at $0.2<z<1$ in forthcoming papers \citep[e.g.,][]{hiner2012}.

\acknowledgments
We are grateful to the referee for a very detailed report that helped improved the contents and the presentation of this paper.
We thank Barry Rothberg for providing a copy of his manuscript before publication and Chien Peng for insightful discussions and suggestions.   We also thank Alan Stockton and Michael Harrison for their assistance in obtaining some of the observations, and Roozbeh Davari for providing scripts for radial plots.  
Support for this program was provided by the 
National Science Foundation, under grant number AST 0507450. Additional support was provided by NASA through a grant from the Space Telescope Science Institute (Program AR-12626), which is operated by the Association of Universities for Research in Astronomy, Incorporated, under NASA contract NAS5-26555, and by the Norwegian Research Council, project number 191735.  M.W. acknowledges funding from the Marie Curie Intra-European Fellowship. 
Some of the data presented herein were obtained at the W.M. Keck Observatory, 
which is operated as a scientific partnership among the California Institute 
of Technology, the University of California and the National Aeronautics and 
Space Administration. The Observatory was made possible by the generous 
financial support of the W.M. Keck Foundation.
The authors wish to recognize and acknowledge the very significant cultural role and reverence that the summit of Mauna Kea has always had within the indigenous Hawaiian community.  We are most fortunate to have the opportunity to conduct observations from this mountain. The National Radio Astronomy Observatory is a facility of the National Science Foundation operated under cooperative agreement by Associated Universities, Inc. 

{\it Facilities:} \facility{Keck:II (ESI)}, \facility{HST (WFPC2)} 

\bibliographystyle{style}
\bibliography{bibsigma}

\end{document}